\documentclass[twocolumn,dvipsnames]{aastex631}

\usepackage{xcolor}
\usepackage{amsmath}
\usepackage{amssymb}
\usepackage{xspace}
\usepackage{enumerate}
\usepackage{acronym}
\usepackage{placeins}
\usepackage[noabbrev,capitalise]{cleveref}

\definecolor{chmagenta}{rgb}{0.54, 0.17, 0.88}

\newcommand\mchirp{\ensuremath{\mathcal{M}_\mathrm{c}}\xspace}
\newcommand\q{\ensuremath{q}\xspace}
\newcommand\chieff{\ensuremath{\chi_\mathrm{eff}}\xspace}
\newcommand\z{\ensuremath{z}\xspace}

\newcommand\chib{\ensuremath{\chi_\mathrm{b}}\xspace}
\newcommand\alphaCE{\ensuremath{\alpha_\mathrm{CE}}\xspace}

\newcommand\betadet{\ensuremath{\beta^{\mathrm{det}}}\xspace}
\newcommand\betadetCE{\ensuremath{\beta^{\mathrm{det}}_\mathrm{CE}}\xspace}
\newcommand\betadetCHE{\ensuremath{\beta^{\mathrm{det}}_\mathrm{CHE}}\xspace}
\newcommand\betadetSMT{\ensuremath{\beta^{\mathrm{det}}_\mathrm{SMT}}\xspace}
\newcommand\betadetGC{\ensuremath{\beta^{\mathrm{det}}_\mathrm{GC}}\xspace}
\newcommand\betadetNSC{\ensuremath{\beta^{\mathrm{det}}_\mathrm{NSC}}\xspace}

\newcommand\betaCE{\ensuremath{\beta_\mathrm{CE}}\xspace}
\newcommand\betaCHE{\ensuremath{\beta_\mathrm{CHE}}\xspace}
\newcommand\betaSMT{\ensuremath{\beta_\mathrm{SMT}}\xspace}
\newcommand\betaGC{\ensuremath{\beta_\mathrm{GC}}\xspace}
\newcommand\betaNSC{\ensuremath{\beta_\mathrm{NSC}}\xspace}

\newcommand\Nobs{\ensuremath{N_\mathrm{obs}}\xspace}
\newcommand\BF[2]{\ensuremath{\mathrm{BF}^{#1}_{#2}}\xspace}

\def\BetaCEGWTC{\ensuremath{90_{-11}^{
+5.0}\%}\xspace}
\def\BetaCHEGWTC{\ensuremath{0.6_{-0.5}^{+1.0}\%}\xspace}
\def\BetaGCGWTC{\ensuremath{3.7_{-3.4}^{+7.6}\%}\xspace}
\def\BetaNSCGWTC{\ensuremath{1.1_{-0.8}^{+1.7}\%}\xspace}
\def\BetaSMTGWTC{\ensuremath{4.0_{-3.2}^{+6.5}\%}\xspace}

\def\BetadetCEGWTC{\ensuremath{21_{-10}^{
+16}\%}\xspace}
\def\BetadetCHEGWTC{\ensuremath{6.9_{-5.2}^{+8.5}\%}\xspace}
\def\BetadetGCGWTC{\ensuremath{24_{-21}^{+26}\%}\xspace}
\def\BetadetNSCGWTC{\ensuremath{20_{-13}^{+17}\%}\xspace}
\def\BetadetSMTGWTC{\ensuremath{24_{-19}^{+24}\%}\xspace}

\def\BetaCEGWTCupper{\ensuremath{95\%}\xspace}
\def\BetadetCEGWTCupper{\ensuremath{37\%}\xspace}

\def\BFalphaCEhighGWTC{\ensuremath{249\xspace}}
\def\BFalphaCElowGWTC{\ensuremath{5 \times 10^{-5}\xspace}}
\def\BFchibGWTC{\ensuremath{1.18\xspace}}

\def\BetaCEOne{\ensuremath{28_{-21}^{
+26}\%}\xspace}
\def\BetaCHEOne{\ensuremath{3.0_{-1.5}^{+2.2}\%}\xspace}
\def\BetaGCOne{\ensuremath{40_{-16}^{+18}\%}\xspace}
\def\BetaNSCOne{\ensuremath{3.6_{-2.1}^{+3.1}\%}\xspace}
\def\BetaSMTOne{\ensuremath{24_{-13}^{+16}\%}\xspace}

\def\BetaCEOneTrue{\ensuremath{40\%}\xspace}
\def\BetaCHEOneTrue{\ensuremath{5\%}\xspace}
\def\BetaGCOneTrue{\ensuremath{25\%}\xspace}
\def\BetaNSCOneTrue{\ensuremath{5\%}\xspace}
\def\BetaSMTOneTrue{\ensuremath{25\%}\xspace}

\def\BetaCESingle{\ensuremath{94\%}\xspace}
\def\BetaCHESingle{\ensuremath{62\%}\xspace}
\def\BetaGCSingle{\ensuremath{41\%}\xspace}
\def\BetaNSCSingle{\ensuremath{31\%}\xspace}
\def\BetaSMTSingle{\ensuremath{51\%}\xspace}

\acrodef{GW}{Gravitational wave}
\acrodef{BBH}{binary black hole}
\acrodef{LVK}{LIGO-Virgo-KAGRA collaboration}
\acrodef{O3}{third observing run}
\acrodef{O3b}{second half of the 3rd observing run}
\acrodef{KDE}{kernel density estimate}

\acrodef{CE}{common envelope}
\acrodef{CHE}{chemically homogeneous evolution}
\acrodef{GC}{globular cluster}
\acrodef{NSC}{nuclear star cluster}
\acrodef{SMT}{stable mass transfer}

\begin{document}

\title{
What You Don't Know Can Hurt You: \\
Use and Abuse of Astrophysical Models in Gravitational-wave Population Analyses}

\author[0009-0007-8996-0735]{April Qiu Cheng}
\affiliation{LIGO Laboratory, Massachusetts Institute of Technology, 185 Albany St, Cambridge, MA 02139, USA}
\affiliation{Department of Physics and Kavli Institute for Astrophysics and Space Research, Massachusetts Institute of Technology, 77 Massachusetts Ave, Cambridge, MA 02139, USA}
\correspondingauthor{April Qiu Cheng}
\email{aqc@mit.edu}
\author[0000-0002-0147-0835]{Michael Zevin}\thanks{NASA Hubble Fellow}
\affiliation{Kavli Institute for Cosmological Physics, The University of Chicago, 5640 South Ellis Avenue, Chicago, IL 60637, USA}
\affiliation{Enrico Fermi Institute, The University of Chicago, 933 East 56th Street, Chicago, IL 60637, USA}

\author[0000-0003-2700-0767]{Salvatore Vitale}
\affiliation{LIGO Laboratory, Massachusetts Institute of Technology, 185 Albany St, Cambridge, MA 02139, USA}
\affiliation{Department of Physics and Kavli Institute for Astrophysics and Space Research, Massachusetts Institute of Technology, 77 Massachusetts Ave, Cambridge, MA 02139, USA}

\begin{abstract}

One of the goals of gravitational-wave astrophysics is to infer the number and properties of the formation channels of \acp{BBH}; to do so, one must be able to connect various models with the data. We explore benefits and potential issues with analyses using models informed by population synthesis. We consider 5 possible formation channels of \acp{BBH}, as in ~\cite{2021ApJ...910..152Z}. First, we confirm with the GWTC-3 catalog what ~\cite{2021ApJ...910..152Z} found in the GWTC-2 catalog, i.e. that the data are not consistent with the totality of observed \acp{BBH} forming in any single channel. Next, using simulated detections, we show that the uncertainties in the estimation of the branching ratios can shrink by up to a factor of $\sim 1.7$ as the catalog size increases from 50 to 250, within the expected number of \ac{BBH} detections in LIGO-Virgo-KAGRA’s fourth observing run. Finally, we show that this type of analysis is prone to significant biases. By simulating universes where all sources originate from a single channel, we show that the influence of the Bayesian prior can make it challenging to conclude that one channel produces all signals. Furthermore, by simulating universes where all 5 channels contribute but only a subset of channels are used in the analysis, we show that biases in the branching ratios can be as large as $\sim 50\%$ with 250 detections. This suggests that caution should be used when interpreting the results of analyses based on strongly modeled astrophysical sub-populations. 

\end{abstract}


\section{Introduction} \label{sec:intro}

\acp{GW} emitted by the mergers of compact objects, neutron stars and black holes, encode the properties of their sources including masses, spins, and distance. 
When one has a large enough dataset, information can be combined from the detected sources to infer properties of the underlying astrophysical process --- or processes --- that created them. 
Nearly 100 compact binary coalescences\footnote{The exact number depends on how conservative of a detection threshold one uses.} (the large majority of which are \acp{BBH}) have been revealed in the data of ground-based GW detectors, LIGO~\citep{2015CQGra..32g4001L} and Virgo~\citep{2015CQGra..32b4001A}, up to their \ac{O3}~\citep{2021arXiv211103606T,2022PhRvD.106d3009O,2023ApJ...946...59N}, allowing this type of analysis to be performed. 
The formation scenarios for compact binaries can be broadly separated in two categories: isolated evolution in the galactic field and dynamical assembly in dense environments such as clusters and AGN disks~\citep[see e.g.][for reviews]{2021hgwa.bookE..16M,2022PhR...955....1M}. 
The overall dataset might contain \acp{BBH} formed from a combination of these and other evolutionary pathways. 

Ideally, one would like to fully characterize any astrophysical formation channels that contribute sources, as well as their relative abundances (branching {ratios}). 
In practice, several approaches have been proposed and followed in the literature. 
They all have merits and shortcomings, and we quickly review them here, focusing on \acp{BBH}, which are the topic of our work. 

\begin{itemize}
    \item \textit{Heuristic models --- } The most straightforward analyses rely on heuristic parametric distributions to describe the astrophysical distributions of black hole parameters. 
    For example, the primary (i.e. most massive) black hole mass distribution can be modeled as a mixture model of a power law and a gaussian (the \texttt{power law + peak} model of ~\citealt{2023PhRvX..13a1048A}, originally introduced by ~\citealt{2018ApJ...856..173T}); the spin orientation distribution as a mixture model of an isotropic component and a component nearly aligned with the orbital angular momentum~\citep{2017PhRvD..96b3012T,2017CQGra..34cLT01V}; etc. 
    The functional forms might be chosen based on computational expediency, or be inspired by reasonable astrophysical expectations (e.g. a power-law component in the black hole mass function because the masses of progenitor stars are distributed that way~\citep{2001MNRAS.322..231K}). 
    Meanwhile, mixture models might allow for different sub-populations to be accounted for. 
    The main potential shortfall of this approach is that if the models are very strong, the resulting posteriors might actually be model-driven, especially for hard-to-measure parameters. 
    This has been shown, for example, in the context of the spin magnitude measurement~\citep{2022ApJ...937L..13C,2021ApJ...921L..15G,2021PhRvD.104h3010R} and the spin orientation~\citep{2022A&A...668L...2V}. 
    On the positive side, if a parameter can be reliably measured, it often has a clear connection with a meaningful astrophysical quantity (e.g., the slope of a mass power law). 
    Heuristic models also allow correlations between parameters to be probed in a straightforward manner~\citep{2021ApJ...922L...5C,2022ApJ...932L..19B,2022A&A...668L...2V,2022MNRAS.517.3928A,2023ApJ...946...50B}. 
    \item \textit{Flexible models ---} Flexible model approaches have been proposed as a way of avoiding the risk of forcing features into the data. 
    In such models, 1D posterior distributions (usually fully marginalized, e.g. $p(m_1 | d)$), are modeled as splines, Gaussian processes, or using autoregression~\citep{2017MNRAS.465.3254M,2019ApJ...886L...1V,2021CQGra..38o5007T,2021ApJ...922..258V, 2022ApJ...928..155T,2022MNRAS.509.5454R,2022ApJ...924..101E,2023ApJ...946...16E,2023arXiv230207289C,2022arXiv221012287G}. 
    While these approaches are less likely to impose stringent features into the posteriors (though see e.g. \citealt{2023arXiv230100834F}) the number of unknown parameters is typically larger than for heuristic models and the model parameters are not usually associated to any specific astrophysical quantity. 
    Furthermore, these methods are not well suited to disentangle sub-populations, and are instead better suited to measuring the overall distribution of parameters. 
    \item \textit{Astrophysically informed models --- } Finally, one may use models obtained directly from the output of population synthesis. 
    Given a set of initial conditions and choices for uncertain stellar, binary, and environmental physical parameters, such models make predictions for the anticipated underlying and detectable distribution of compact binaries. 
    These models can be parametrized directly in terms of physically-meaningful quantities (e.g. the onset and evolution of binary mass transfer phases, the strength of supernova natal kicks, the efficiency of angular momentum transport), and correlations between parameters are automatically built in the models, both of which are very strong positive factors. 
    In practice, the range of variations in physical uncertainties, as well as the number and complexity of the differing channels one considers, are often limited by the availability of numerical simulations that thoroughly explore the variation of the output (e.g. distribution of spin magnitude in the population) when one of the physical input parameters (e.g. efficiency of common envelope evolution) is varied. 
    This is due to the fact that most population synthesis algorithms require significant computational resources to run, which implies that they cannot be evaluated ``on the fly'' for any value of their input parameters but must instead, for example, be evaluated on a sparse grid. 
    Recent efforts have begun to more thoroughly explore compact binary population predictions for individual formation channels over an expansive array of physical and environmental uncertainties~\citep[e.g.,][]{2022MNRAS.516.5737B}. 
    Consideration of multiple formation channels simultaneously and self-consistently proves more difficult given the diversity of codebases needed to model different channels and the unique physics that affects compact binary populations from each channel. 
    The most expansive multi-channel analysis to date was performed in~\cite{2021ApJ...910..152Z}, who considered 5 possible formation channels (see~Section \ref{subsec:inferencemethod} below), parameterized by the spin of quasi-isolated black holes at birth (a proxy for the efficiency of angular momentum transport in massive stars) and the efficiency of the common envelope ejection. 
    By analyzing the 45 confident \ac{BBH} sources of the penultimate (GWTC-2) \ac{LVK} catalog, ~\cite{2021ApJ...910..152Z} found that the data required more than a single formation channel in order to explain the diversity of \ac{GW} events and the distribution of parameters of the detected binaries. 
    They also found that the data preferred small natal black hole spins, consistent with the fact that most of the \ac{LVK} \acp{BBH} have small spin magnitudes. 
\end{itemize}
In this paper, we analyze both merits and shortcoming of approaches based on models informed by the output of population synthesis codes. 
First, we repeat the analysis of ~\cite{2021ApJ...910..152Z} on the latest \ac{LVK} catalog (GWTC-3), which comprises 69 \ac{BBH} with a false alarm rate of less than $1~\mathrm{yr}^{-1}$. 
We find that the model for common envelope evolution can explain up to \BetaCEGWTCupper of the \acp{BBH} in the underlying population (while contributing up to \BetadetCEGWTCupper of the \textit{detectable} \acp{BBH}). 

Then, we create catalogs of simulated \ac{BBH} signals with parameters drawn from our models, for some assumed values of the common envelope efficiency, quasi-isolated natal black hole spins, and branching fractions across channels. 
We analyze the performance of the analysis as the number of \ac{BBH} sources in the catalog increases from 50 to 250. 
We find that for channels that produce higher number of detectable sources (and hence are proportionally more represented in the catalog, even if the true underlying branching fractions are not higher) the uncertainty on the underlying branching fraction can improve by a factor of up to $\sim 1.7$ as the number of sources increases to 250. 

Next, we study the biases that can be introduced in this type of inference if the analysis is not using a suite of models fully representative of what is actually realized in nature. 
This problem was first indirectly shown by \cite{2022PhRvD.105h3526F}, who added a channel for primordial black hole formation to those used by ~\cite{2021ApJ...910..152Z}, and obtained that the inference on the fraction of primordial back holes was significantly affected by which of the \textit{other} channels were included in the analysis. 
To do so, we generate mock universes where each of the 5 channels contributes some known fraction of the underlying population, and run the inference excluding in turn one of the 5 models. 
We show how this introduces biases in the inference of the remaining 4 channels' branching fractions. 
The channels that are most heavily biased are the ones that can most easily produce sources similar to the one channel excluded from the analysis, as well as those with the lowest detection efficiencies. 
Finally, we generate universes where the totality of the \ac{BBH} sources are produced by one of the 5 channels, and run the analysis with the same 5 models. 
We show that while the natal spin can be inferred correctly, it is usually not possible to exclude that more than one channel contributes to the population after 100 events, a caveat to our result from our inference on GWTC-3 data that multiple channels contribute to both the underlying and detected \ac{BBH} population.

The rest of the paper is organized as follows: in Section \ref{sec:methods_and_gwtc3} we review the basics of hierarchical inference and the models used in this work, and apply these tools to GWTC-3's \acp{BBH}. 
In Section \ref{sec:convergence} we apply the method to different simulated catalogs in the ideal scenario where models and true populations match. 
In Section \ref{sec:biases} we focus on biases in the inference. 
We conclude in Section \ref{sec:conclusions}.

\section{Hierarchical Inference on GWTC-3 Data}\label{sec:methods_and_gwtc3}    

\subsection{Hyper-Inference Method}\label{subsec:inferencemethod}

We use hierarchical Bayesian inference on the branching fractions between different astrophysical formation channels of \acp{BBH}. 
Our methods mostly follow the analysis developed by \cite{2021ApJ...910..152Z}, adapting their codebase, Astrophysical Model Analysis and Evidence Evaluation (AMA$\mathcal{Z}$E), for our work. 
Here we outline the essentials of this method, as well as the key differences. 

We consider five different formation channels: three isolated evolution (field) channels, and two dynamical formation channels. 
The \ac{CE} \citep[e.g.][]{1976IAUS...73...35V,1976IAUS...73...75P,1993MNRAS.260..675T,1998ApJ...506..780B,2002ApJ...572..407B,2012ApJ...759...52D,2016MNRAS.462.3302E,2016ApJ...819..108B,2017NatCo...814906S,2018MNRAS.480.2011G} and \ac{SMT} \citep{2017MNRAS.471.4256V,2019MNRAS.490.3740N,2022ApJ...940..184V,2021ApJ...922..110G} scenarios are field channels which involve unstable and stable mass transfer, respectively, following the formation of the first black hole. 
In the \ac{CHE} channel, stars in a close, tidally-locked binary rotate rapidly, causing temperature gradients that lead to efficient mixing of the stars' interiors. 
The stars do not undergo significant expansion of the envelope, preventing significant post-main sequence wind mass loss and premature merging, resulting in higher mass BBHs \citep{2016MNRAS.460.3545D,2016MNRAS.458.2634M,2016A&A...588A..50M}. 
Finally, the two dynamical formation channels lead to merging black holes via strong gravitational encounters that harden the binary in cluster cores \citep[e.g.][]{1991ApJ...372..111M,1992PASP..104..981H,1993ApJ...415..631S,2000ApJ...528L..17P,2002ApJ...576..894M,2006ApJ...637..937O,2006ApJ...640..156G,2007ApJ...658.1047F,2010MNRAS.407.1946D,2014ApJ...784...71S,2014MNRAS.441.3703Z,2015PhRvL.115e1101R,2016PhRvD..93h4029R,2016ApJ...831..187A,2017MNRAS.464L..36A,2017ApJ...840L..14S,2020ApJS..247...48K}, where heavy black holes migrate towards due to dynamical friction \citep{1978RvMP...50..437L,1993Natur.364..423S}; we consider dynamical formation of \acp{BBH} in \acp{GC} and \acp{NSC}. 
See \citet{2021ApJ...910..152Z} for a more detailed description of the astrophysical models considered in this work. 

\begin{figure*}[ht!]
    \includegraphics[width=\textwidth]{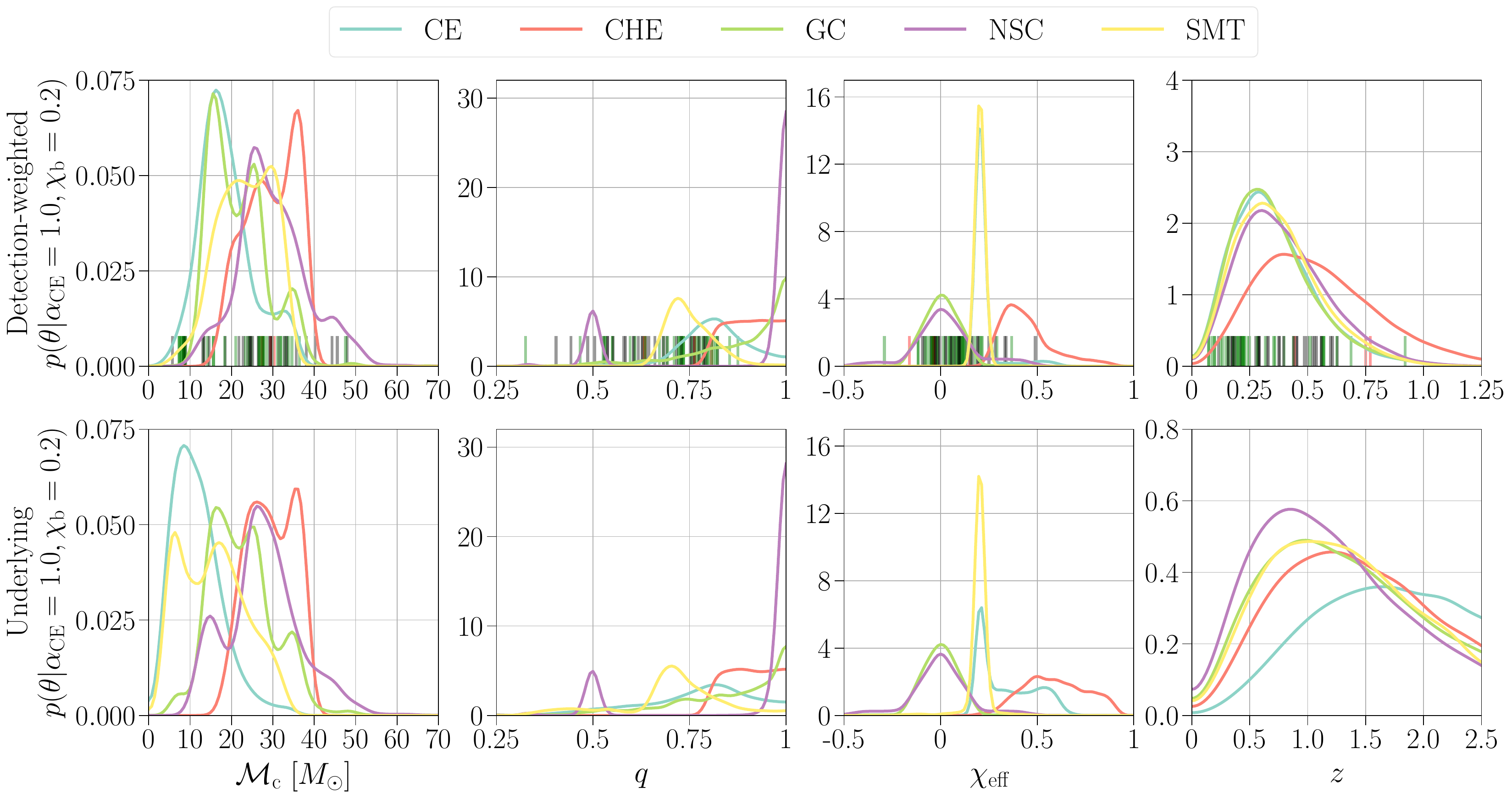}
    \caption{Detection-weighted (top) and underlying (bottom) marginalized \acp{KDE} of our 4 \ac{BBH} parameters, $\vec{\theta} = [\mchirp, \q, \chieff, \z]$, for our 5 different formation channels with $\alphaCE=1.0$, $\chib=0.2$. 
    Note the unique features of each formation channel and their dependence on the chosen model $\chib=0.2$: while the \chieff marginalized \ac{KDE} for field channels \ac{CE} and \ac{SMT} peak at $\chieff=\chib=0.2$, with the \ac{CE} channel having support for larger \chieff through tidal spin-up, dynamical channels \ac{GC} and \ac{NSC} peak at $\chieff=0$ due to the independently isotropically oriented spins of the \acp{BBH} with symmetric wings to more extreme \chieff values from hierarchical mergers, and the \ac{CHE} channel has \chieff exceeding $\chib=0.2$ due to significant tidal spin-up of both black hole progenitors. 
    The first row is the same as the third row of Figure 1 from \cite{2021ApJ...910..152Z}, except here we have calculated the detection weighting with O4 sensitivities, and the black ticks show the median values of the parameter estimation posteriors for each event in GWTC-2, while the green and red ticks mark the events added to and removed from, respectively, this updated analysis that includes GWTC-3. } 
    \label{fig:popmodels}
\end{figure*}

Each of these channels are modelled to predict the 4-dimensional distribution of the \acp{BBH} it forms with parameters ${\vec{\theta} = [\mchirp, \q, \chieff, \z]}$, where \mchirp is the source-frame chirp mass, \q is the mass ratio (defined to be $0 < \q \leq 1$), \chieff is the effective dimensionless spin parameter, and \z is the redshift; these are constructed into a probability distribution using a 4-dimensional \ac{KDE} bounded by the physical constraints of each parameter. 
The models also depend on two additional parameters encoding uncertainties in the physical prescription: \chib, the dimensionless spin of a black hole formed in quasi-isolation directly following core collapse, and \alphaCE, which parameterizes the efficiency of common envelope ejection \citep[see e.g.,][]{2013A&ARv..21...59I}.\footnote{The choice of these hyperparameters come from the availability of self-consistent astrophysical models; in particular, the globular cluster simulations were run on a grid of discrete values in \chib, and \alphaCE variations only affect the parameter distributions of the \ac{CE} model, which was less computational expensive to explore additional variations than other models. See \cite{2021ApJ...910..152Z} for further discussion on the reasoning and motivations for the physical prescriptions considered.} 
We note that the choice of natal black hole spin \chib does not set all black holes in a given population to merge with this exact spin because tidal spin-up processes \citep{2018A&A...616A..28Q,2018MNRAS.473.4174Z,2021A&A...647A.153B} and hierarchical mergers \citep{2005PhRvL..95l1101P,2007PhRvL..98i1101G,2008PhRvD..77b6004B} can increase the spin of black holes that participate in \ac{BBH} mergers. 
We assume these two parameters take on a grid of discrete values ($\chib \in [0.0, 0.1, 0.2, 0.5]$, $\alphaCE \in [0.2, 0.5, 1.0, 2.0, 5.0]$) over which we compute the models, although \alphaCE only affects the \ac{CE} channel in our models.
A plot of the detection-weighted marginalized model \acp{KDE} for $\alphaCE=1.0$, $\chib=0.2$ is shown in \cref{fig:popmodels} as an example. 
Further details, including formation models, detection weighting, and mathematical framework, can be found in \cite{2021ApJ...910..152Z}. 

Overall, we perform hierarchical inference on 7 hyperparameters\footnote{Due to the restriction $\sum_i{\beta_i}=1$ (the branching fractions must add up to $1$), we technically only perform inference on $6$ hyperparameters.} $\vec{\Lambda} = [\vec{\beta}, \chib, \alphaCE]$, where $\vec{\beta} = [\betaCE, \betaCHE, \betaSMT, \betaGC, \betaNSC]$ are the 5 astrophysical formation channel branching fractions. 
The steps involved in perfoming hierarchical inference on GW populations given a set of posterior samples ($\vec{\theta} = [\mchirp, \q, \chieff, \z]$ in our case) for \Nobs sources in the presence of selection effects have been thoroughly discussed in the literature~\citep{2015PhRvD..91b3005F,2019MNRAS.486.1086M,2019PASA...36...10T,2022hgwa.bookE..45V,2023PhRvX..13a1048A}, and we therefore do not review them here. 
As in \cite{2021ApJ...910..152Z}, we use a broad, agnostic prior for the hyperparameters: a flat symmetric Dirichlet prior for $\vec{\beta}$ and a uniform prior for \chib and \alphaCE over the allowed discrete values. 
The outputs of the inference are samples of the hyperposterior $p(\vec{\beta} | \chib, \alphaCE)$ over the grid of hypermodels (i.e. allowed values of \chib and \alphaCE), from which we can compute the marginalized hyperposterior $p(\vec{\beta})$ as well as the Bayes factors \BF{a}{b} of hypermodel $a$ compared to hypermodel $b$. 
Bayes factors $>1$ indicate that model $a$ is more supported than model $b$, with $\BF{a}{b}>10\,(100)$ indicating strong (decisive) support for model $a$ over model $b$ \citep{bayesfactor}.

We also recover the \textit{detectable} branching fractions \betadet, which represent the fraction of detectable \acp{BBH} originating from each channel. 
These are computed by re-scaling each underlying branching fraction by its detection efficiency, defined as $\xi^{\chib, \alphaCE}_j = \int  P_\mathrm{det}(\vec{\theta})p(\vec{\theta}|\mu_j^{\chib, \alphaCE})\,\mathrm{d}\vec{\theta}$, where  $P_\mathrm{det}(\vec{\theta})$ is the probability of detecting a \ac{BBH} with parameters $\vec{\theta}$ and $p(\vec{\theta}|\mu_j^{\chi, \alpha})$ is the probability of formation channel $j$ producing a \ac{BBH} with parameters $\vec{\theta}$, dependent on the model $\mu_j$ and choice of physical prescription \chib and \alphaCE \citep{2021ApJ...910..152Z}.
We apply this method to both real (Section \ref{subsec:application_to_gwtc3}) and simulated (Sections \ref{sec:convergence}, \ref{sec:biases}) \ac{BBH} data.

\subsection{Application to GWTC-3} \label{subsec:application_to_gwtc3}

\begin{figure*}[ht!]
    \includegraphics[width=\textwidth]{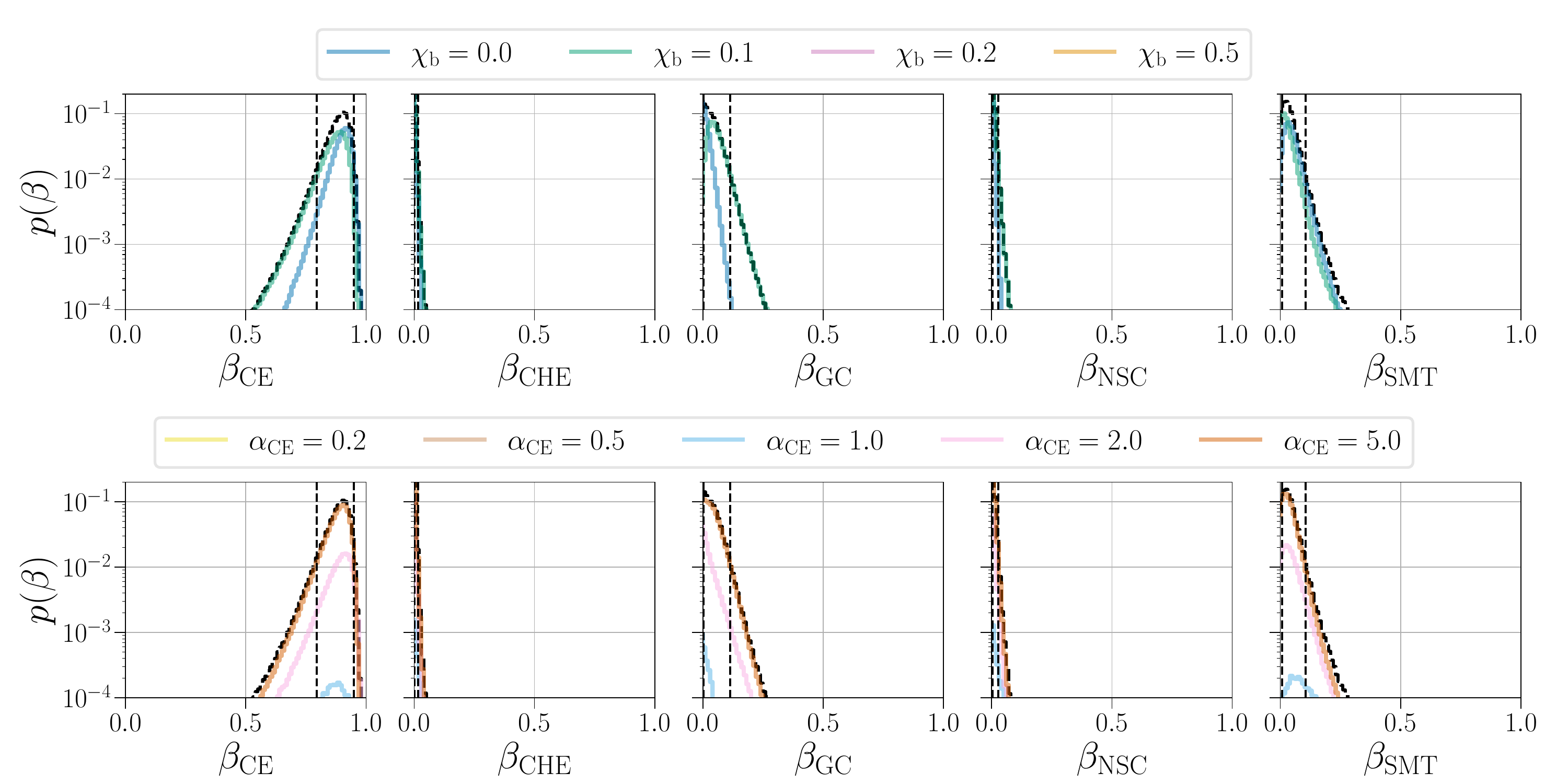} 
    \caption{Marginalized branching fractions inferred from 68 \acp{BBH} from the cumulative GWTC-3 for each of the five formation channels. 
    Colored curves show the contributions from different values of \chib (top row) and \alphaCE (bottom row), marginalized over \alphaCE and \chib, respectively; the black dotted curve shows the total contribution from all hypermodels. 
    Note the log-scaling of the y-axis; the 90\% symmetric credible intervals are marked by vertical lines.}
    \label{fig:gwtc3}
\end{figure*}

We extend the work of \cite{2021ApJ...910..152Z} by applying the inference to confident \ac{BBH} detections up to the \ac{O3b}. 
We use the publicly-released priors and posterior samples from the GWTC-2.1 \citep{2021arXiv210801045T} and GWTC-3 \citep{2021arXiv211103606T} analyses; detection probabilities are calculated for the LIGO-Hanford, LIGO-Livingston, and VIRGO network operating at \texttt{midhighlatelow} sensitivities \citep{2018LRR....21....3A}. 
There are two key differences between our analysis and that of  \cite{2021ApJ...910..152Z}. 
First, we apply a more stringent detection threshold of $\text{FAR} \leq 1~\text{yr}^{-1}$ \citep{2023PhRvX..13a1048A}; therefore, events GW190424\_180648, GW190514\_065416, and GW190909\_114149, which were used in the previous analysis, are now excluded. 
Second, we evaluate the prior at the posterior points analytically rather than by constructing a \ac{KDE} from prior samples; for further discussion on this point see Appendix~\ref{app:ptheta}. 
Finally, as in \cite{2021ApJ...910..152Z}, we opt to exclude GW190521 and GW190814 from the analysis, as their posteriors extend significantly to regions of our model \acp{KDE} with little to no support. 
Furthermore, in the case of GW190814, it is uncertain whether this system is comprised of two black holes or a neutron star and a black hole.
We find that the inclusion of GW190521 does not significantly affect our results; see \cite{2022PhRvD.105h3526F} for an analysis that includes GW190521. 
Overall, we do hierarchical inference on 68 \acp{BBH}, compared to the 45 \acp{BBH} considered in \cite{2021ApJ...910..152Z}. 
Unless otherwise specified, we report results as median and 90\% symmetric credible intervals. 

\cref{fig:gwtc3} shows the posterior distributions on the underlying branching fraction $\vec{\beta}$ including detections from \ac{O3b}; the same plot but for the detectable branching fractions can be found in \cref{fig:gwtc3-det} in Appendix~\ref{app:additionalfigs}. 
We find $\betaCE=\BetaCEGWTC$, $\betaCHE=\BetaCHEGWTC$, $\betaGC=\BetaGCGWTC$, $\betaNSC = \BetaNSCGWTC$ and $\betaSMT = \BetaSMTGWTC$, indicating strong support for the \ac{CE} channel dominating the underlying astrophysical population in our set of models. 
However, there is comparable contribution from all five channels to the \emph{detectable} \ac{BBH} population:  $\betadetCE=\BetadetCEGWTC$, $\betadetCHE=\BetadetCHEGWTC$, $\betadetGC=\BetadetGCGWTC$, $\betadetNSC = \BetadetNSCGWTC$, and $\betadetSMT = \BetadetSMTGWTC$. 
We attribute this difference to the fact that compared to other channels, the \ac{CE} channel produces less massive black holes distributed at higher redshifts, which therefore are harder to detect; see \cref{fig:popmodels}. 
With 90\% (99\%) credibility, no single formation channel contributes to more than 49\% (61\%) of the detectable \ac{BBH} population. 
Additionally, over 98\% of posterior samples have significant ($\betadet>10$\%) contributions from three or more different formation channels. 
Overall, we find that the \ac{CE} channel contributes the most to the underlying \ac{BBH} population; however, as we discuss in Section \ref{sec:biases}, the posterior for \betaCE is also the most uncertain and prior-dominated. We additionally find that a mix of formation channels contributes to the detectable \ac{BBH} population
These results, as well as the overall shapes of the branching fraction posteriors, are consistent with the analysis with GWTC-2 data \citep{2021ApJ...910..152Z}, which found $\betaCE=71^{+19}_{-60}\%$, and that no single channel contributed to more than 70\% of the detectable \ac{BBH} population with 99\% confidence. 
No strong correlations are apparent upon examining a corner plot of the branching fractions.

Turning to the selection of physical prescription hyperparameters \chib and \alphaCE, we find no posterior support for models with $\chib > 0.1$; there is no significant preference between the $\chib=0.0$ and $\chib=0.1$ models, with $\BF{\chib=0.1}{\chib=0.0} = \BFchibGWTC$. 
We favor high common envelope efficiencies, with $\BF{\alphaCE=5.0}{\alphaCE=1.0}=\BFalphaCEhighGWTC$, and strongly disfavor low common envelope efficiencies, with $\BF{\alphaCE=0.2}{\alphaCE=1.0}=\BFalphaCElowGWTC$. 
These results are also consistent with \cite{2021ApJ...910..152Z}. 
The main difference is that we obtain stronger constraints, both in singling out preferred hypermodels and in the uncertainties (widths) of the branching fraction hyperposteriors. 
We attribute this effect to the increase in sample size from GWTC-2 to GWTC-3, as well as the difference in method in evaluating the prior at the posterior points during the inference; see Appendix~\ref{app:ptheta} for further discussion. 

\section{Projections for Future Catalogs} \label{sec:convergence}

\subsection{Method} \label{subsec:injsmethod}

\begin{figure*}[ht]
    \includegraphics[width=\textwidth]{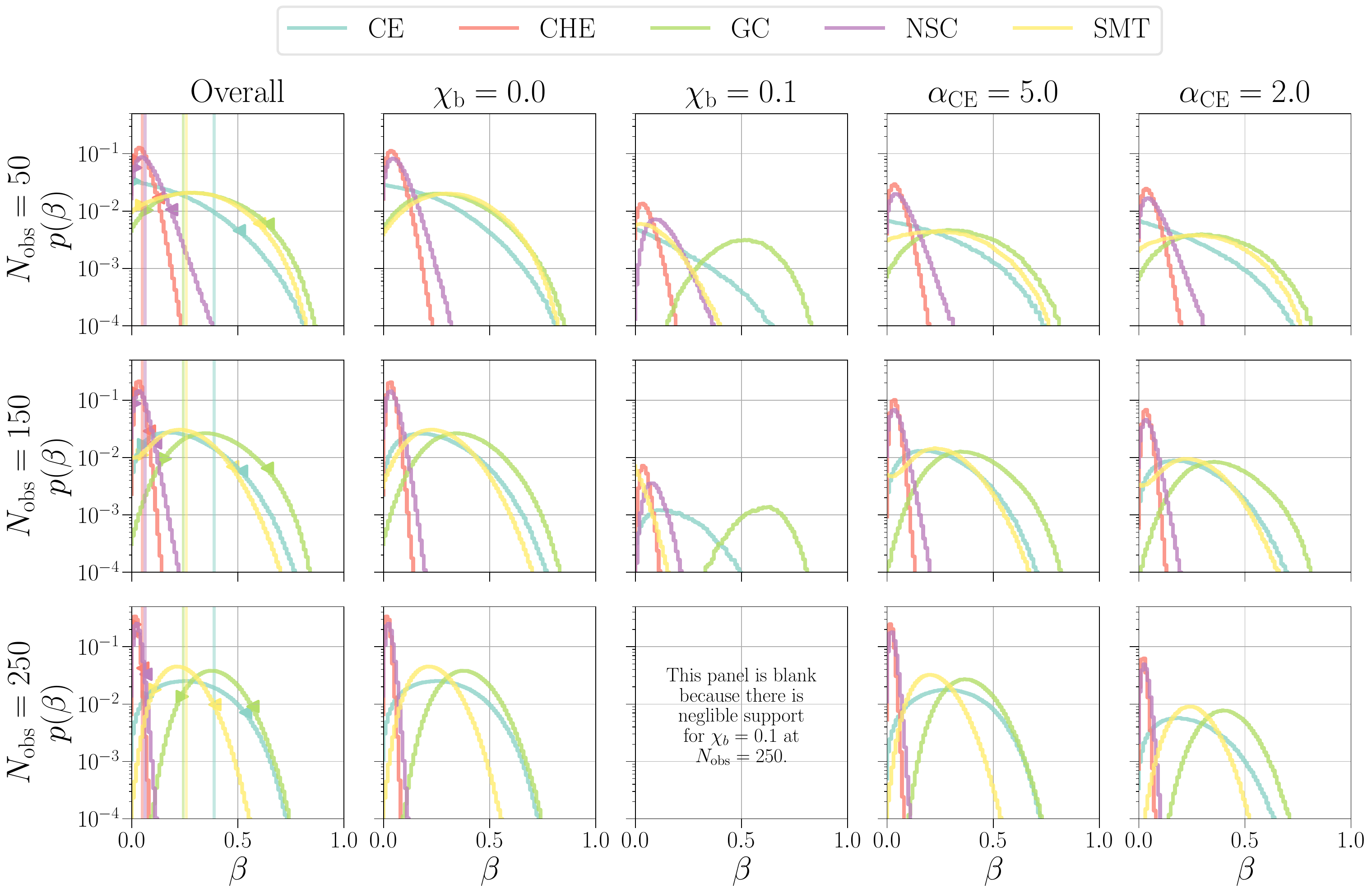}
    \caption{Marginalized branching fraction posteriors for 50 (first row), 150 (second row) and 250 (third row) mock detections. 
    Here, branching fraction posteriors of different channels are plotted together as different colored lines, while contributions from different hyper-models are shown in different columns. 
    The first column (from the left) shows the total marginalized branching fraction; the triangles mark the credible symmetric 90\% intervals. 
    The second and third columns show the contributions from the two values of \chib (marginalized over \alphaCE) with the most support; the fourth and fifth columns are analogous for \alphaCE. 
    Note that the second through fifth columns do \textit{not} represent normalized probability distributions; we plot them to scale with their contribution to the total marginalized posterior in the first column, just as with the colored curves in \cref{fig:gwtc3}. 
    True values of the underlying branching fractions ($\betaCE=\BetaCEOneTrue$, $\betaCHE=\BetaCHEOneTrue$, $\betaGC=\BetaGCOneTrue$, $\betaNSC=\BetaNSCOneTrue$, $\betaSMT=\BetaSMTOneTrue$) are shown by the vertical colored lines in the ``Overall'' column, where \betaNSC and \betaSMT are given small artificial offsets, while $\chib=0.0$ and $\alphaCE=1.0$. }
    \label{fig:convergence1}
\end{figure*}

Next, we perform the same inference using simulated \ac{BBH} observations in a universe where the \ac{BBH} population exactly follows our models. 
The motivation for this analysis is to to quantify how uncertainties in the hyperposterior scale with the number of observed events. 

We first create a mock ``universe'', i.e. a set of true values for our hyperparameters $\vec{\Lambda} = [\vec{\beta}, \chib, \alphaCE]$. 
From each formation channel $j$, we draw $n_j = \beta_j n$ \acp{BBH} from the population model $p(\theta | \mu_j^{\chib,\, \alphaCE})$, for a total of $n=5 \times 10^4$ \acp{BBH} that form our underlying population. 
Next, we draw from this underlying population, assigning extrinsic parameters (sky location and inclination) from an isotropic distribution. 
For each \ac{BBH} system, we calculate its optimal signal-to-noise ratio $\rho_\mathrm{opt}$ assuming a network consisting of LIGO-Hanford, LIGO-Livingston, and Virgo operating at O4 low (LIGO) and high (Virgo) sensitivities \citep{2018LRR....21....3A, DCC_T2000012}, and keep only the mock signals with $\rho_\mathrm{opt} \geq 11$; we repeat this process until we have a mock catalog of $\Nobs$ detections.
Then, we perform parameter estimation on the \Nobs \acp{BBH}. 
For both the SNR calculation and parameter estimation, we use the Bayesian inference software \texttt{bilby} \citep{2019ApJS..241...27A} and the \texttt{IMRPhenomXP} waveform approximant \citep{2021PhRvD.103j4056P}. 
Finally, we use these posterior samples for the hierarchical inference analysis outlined in Section \ref{subsec:inferencemethod}. 
For consistency, we use the same O4 sensitivities for the detection weighting during the inference as the above computation of mock signal SNRs. 

\begin{table}[b]
\noindent\makebox[6cm]{
 \begin{tabular}{c|c|ccccccc} 
 Sections & \Nobs & \betaCE & \betaCHE & \betaGC & \betaNSC & \betaSMT \\ 
 \hline
 \ref{subsec:convergenceresults}, \ref{subsec:mixture} & up to 250 & 40\% & 5\% & 25\% & 5\% & 25\% \\
 \ref{subsec:convergenceresults}, \ref{subsec:spinbias}, \ref{subsec:mixture} & up to 250 & 20\% & 20\% & 20\% & 20\% & 20\% \\
 \ref{subsec:singlechannel} & 100 & \multicolumn{5}{c}{$\beta$=100\% for some channel} \\
 \end{tabular}
 }
 \caption{True values of the branching fractions for the different universes from which we generate our mock catalogs, along with the number of detections in each mock catalog used in the inference and the sections in which they are discussed. For all universes we pick true values $\chib=0.0$ and $\alphaCE=1.0$, except the equal branching fraction universe (second row), where we use $\chib=0.2$.}
 \label{tab:universes}
\end{table}

The various mock universes that we use in this paper, along with the sections in which they are discussed, are summarized in \cref{tab:universes}. 
For this analysis, we choose a fiducial unequal mixture of formation channels with underlying branching fractions $\betaCE=\BetaCEOneTrue$, $\betaCHE=\BetaCHEOneTrue$, $\betaGC=\BetaGCOneTrue$, $\betaNSC=\BetaNSCOneTrue$, $\betaSMT=\BetaSMTOneTrue$, quasi-isolated natal spins of $\chib = 0.0$, and a \ac{CE} efficiency of $\alphaCE = 1.0$. 
To investigate the scaling of the hyperposterior with \Nobs, for each universe we repeat the inference with $\Nobs=50$, $\Nobs=150$, and $\Nobs=250$. 
$\Nobs=250$ represents an estimate on the total number of \ac{BBH} detections anticipated by the end of O4 \citep{2023arXiv230609234W}. 

\subsection{Results}\label{subsec:convergenceresults}

\begin{figure*}[ht]
    \includegraphics[width=\textwidth]{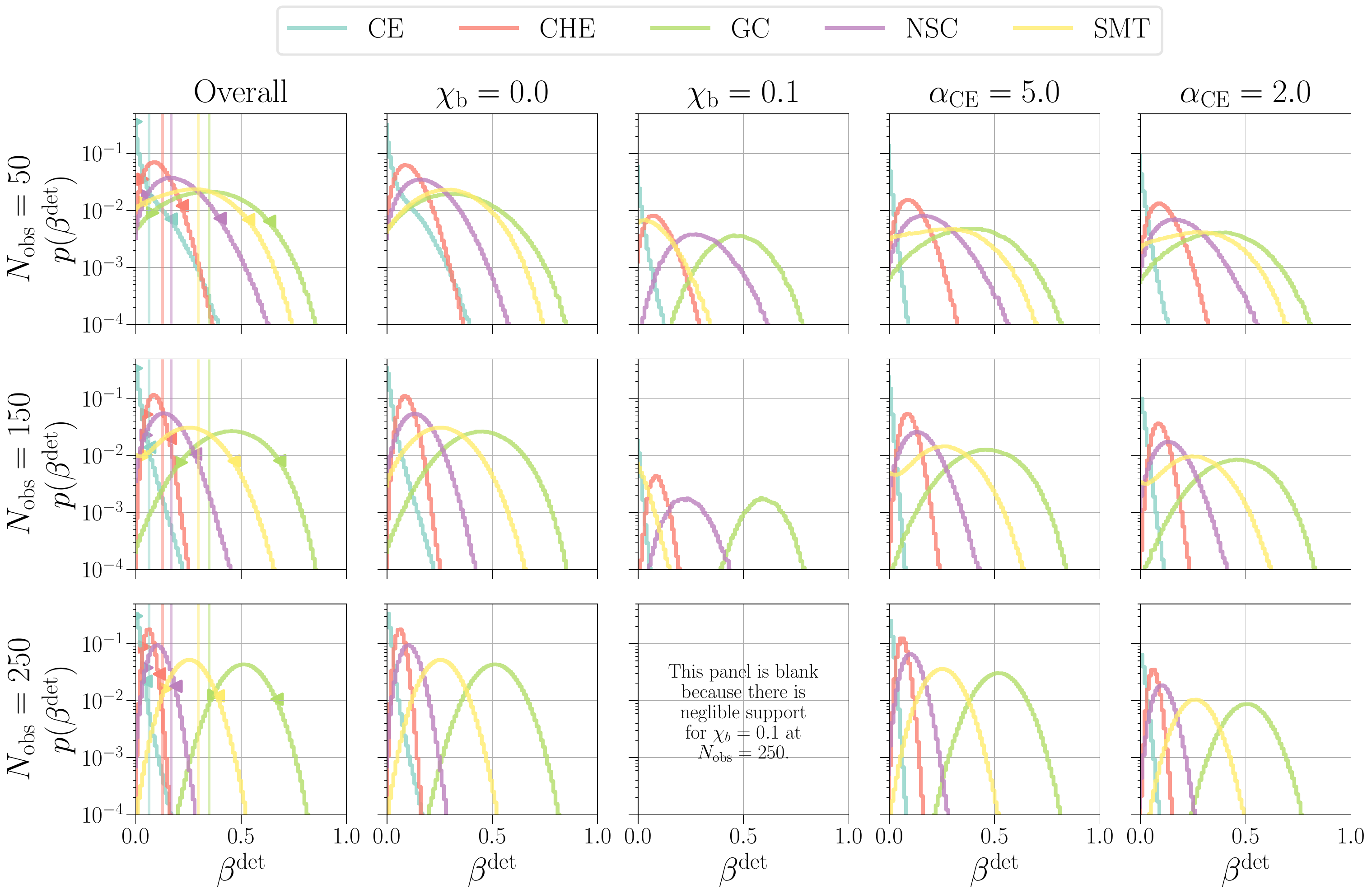}
    \caption{The same figure as \cref{fig:convergence1} except with detectable branching fractions rather than underlying. }
    \label{fig:convergence1-det}
\end{figure*}

In \cref{fig:convergence1}, we plot for our chosen fiducial universe the overall posterior distribution on the underlying branching fractions for different values of \Nobs (left column), as well as contributions (to scale) from the most favored values of \chib and \alphaCE (columns 2 to 5). 
First, we do recover the true values of $\vec{\beta}$ and \chib. 
For $\Nobs=250$, we find $\betaCE=\BetaCEOne$, $\betaCHE=\BetaCHEOne$, $\betaGC=\BetaGCOne$, $\betaNSC = \BetaNSCOne$ and $\betaSMT = \BetaSMTOne$; the true value of the branching fraction falls within the 90\% symmetric credible interval of the posterior for all 5 channels. 
Furthermore, $\chib=0$, the chosen true value for \chib, is favored over the next best model, $\chib=0.1$, by a Bayes factor of $3.5 \times 10^5$. 

Unlike \chib, the model with $\alphaCE$ equal to the true value is not favored, with a marginal preference for higher values $\alphaCE=5.0$ and $\alphaCE=2.0$ over the true value $\alphaCE=1.0$ with Bayes factors $\BF{\alphaCE=5.0}{\alphaCE=1.0}=8.9$ and  $\BF{\alphaCE=2.0}{\alphaCE=1.0}=2.5$. 
Since $\alphaCE$ only affects the \ac{CE} channel, it is not too surprising that the inference did not decisively favor any one value, in a universe where most detected \acp{BBH} do not come from the \ac{CE} channel. 
We note two possible contributing factors to favoring higher values of $\alphaCE$ over the true value. 
If \acp{BBH} formed in the \ac{CE} channel disperse their envelopes more efficiently (i.e. have higher values of $\alphaCE$), the resulting \acp{BBH} are
\begin{enumerate}[(a)]
    \item Less massive, leading to lower detection efficiencies and therefore larger measurement uncertainties for this channel. This is because models with larger \alphaCE have less low-mass binaries merging within the \ac{CE} itself, leading to more low-mass \acp{BBH} being able to form and merge~\citep{2021A&A...647A.153B}. 
    \item Lower spinning, due to the post-\ac{CE} separations being wider for higher \alphaCE and therefore less susceptible to tidal spin-up \citep{2018A&A...616A..28Q,2018MNRAS.473.4174Z,2021A&A...647A.153B}; this leads to more detections with \chieff closer to $\chib=0.0$, the chosen fiducial value for this universe. 
    In \chieff space, this is a feature degenerate with dynamical channels such as the \ac{GC} channel, which also produces \acp{BBH} with $\chieff\approx0$ due to the \acp{BBH} produced having isotropic spin orientations. 
\end{enumerate}
This result suggests to view our result in Section \ref{subsec:application_to_gwtc3}, that higher common envelope efficiencies are favored, with some caution. 

\begin{table*}[ht]
\noindent\makebox[14.5cm]{
 \begin{tabular}{c|ccccc|ccccc|cc} 
 \Nobs & $\Delta\betaCE$ & $\Delta\betaCHE$ & $\Delta\betaSMT$ & $\Delta\betaGC$ & $\Delta\betaNSC$ & $\Delta\betadetCE$ & $\Delta\betadetCHE$ & $\Delta\betadetSMT$ & $\Delta\betadetGC$ & $\Delta\betadetNSC$ & \BF{\chib=0.0}{\chib=0.1} & \BF{\alphaCE=1.0}{\alphaCE=5.0} \\ 
 \hline
 50 & 49\% & 11\% & 52\% & 17\% & 56\% & 17\% & 19\% & 55\% & 34\% & 49\% & 8.0 & 0.81 \\
 150 & 46\% & 6.2\% & 48\% & 9.7\% & 43\% & 6.4\% & 11\% & 47\% & 24\% & 42\% & 27 & 0.36 \\
 250 & 46\% & 3.7\% & 33\% & 5.2\% & 28\% & 4.6\% & 7.2\% & 30\% & 14\% & 25\% & $3.5 \times 10^5$ & 0.11 \\
 \hline
 \% change & 5.9\% & 66\% & 36\% & 69\% & 49\% & 73\% & 62\% & 46\% & 60\% & 49\% & &
 \end{tabular}
 }
 \caption{Uncertainties in the measurement of the branching fractions  with \Nobs shown in \cref{fig:convergence1,fig:convergence1-det}. 
 The right-most two columns give the Bayes factors of the true value of \chib and \alphaCE over the most favored competing value for 50, 150, and 250 mock detections (first, second, and third row). 
 The rest of the table shows the underlying and detectable branching fraction uncertainties, as well as their percent changes between the first and third rows. 
 Note that these are credible 90\% interval widths in percentage points, not percent uncertainties. }
 \label{tab:convergence1}
\end{table*}

More notably, we see convergence of the hyperposteriors towards their true values as we increase the number of detections. 
First, we highlight the narrowing of the branching fraction posteriors from the first row to the third row of \cref{fig:convergence1}.  
We quantify uncertainties in the branching fraction posteriors by the widths of the 90\% symmetric credible intervals, and will hereafter use the two phrases interchangeably.
These uncertainties decrease by up to 69\% as we go from 50 to 250 mock events (see \cref{tab:convergence1}). 
The Bayes Factor in favor of the correct \chib increases by nearly 5 orders of magnitude, strongly selecting the true value $\chib=0.0$. 
This is illustrated by the empty plot corresponding to the contribution from $\chib=0.1$ for $\Nobs=250$ (third row, middle column), indicating negligible support for competing values of \chib. 
On the other hand, the inference has increasing support for $\alphaCE=5.0$, but even at 250 mock detections, we only weakly prefer it over the true model $\alphaCE=1.0$.  

\begin{figure}[t]
    \includegraphics[width=\columnwidth]{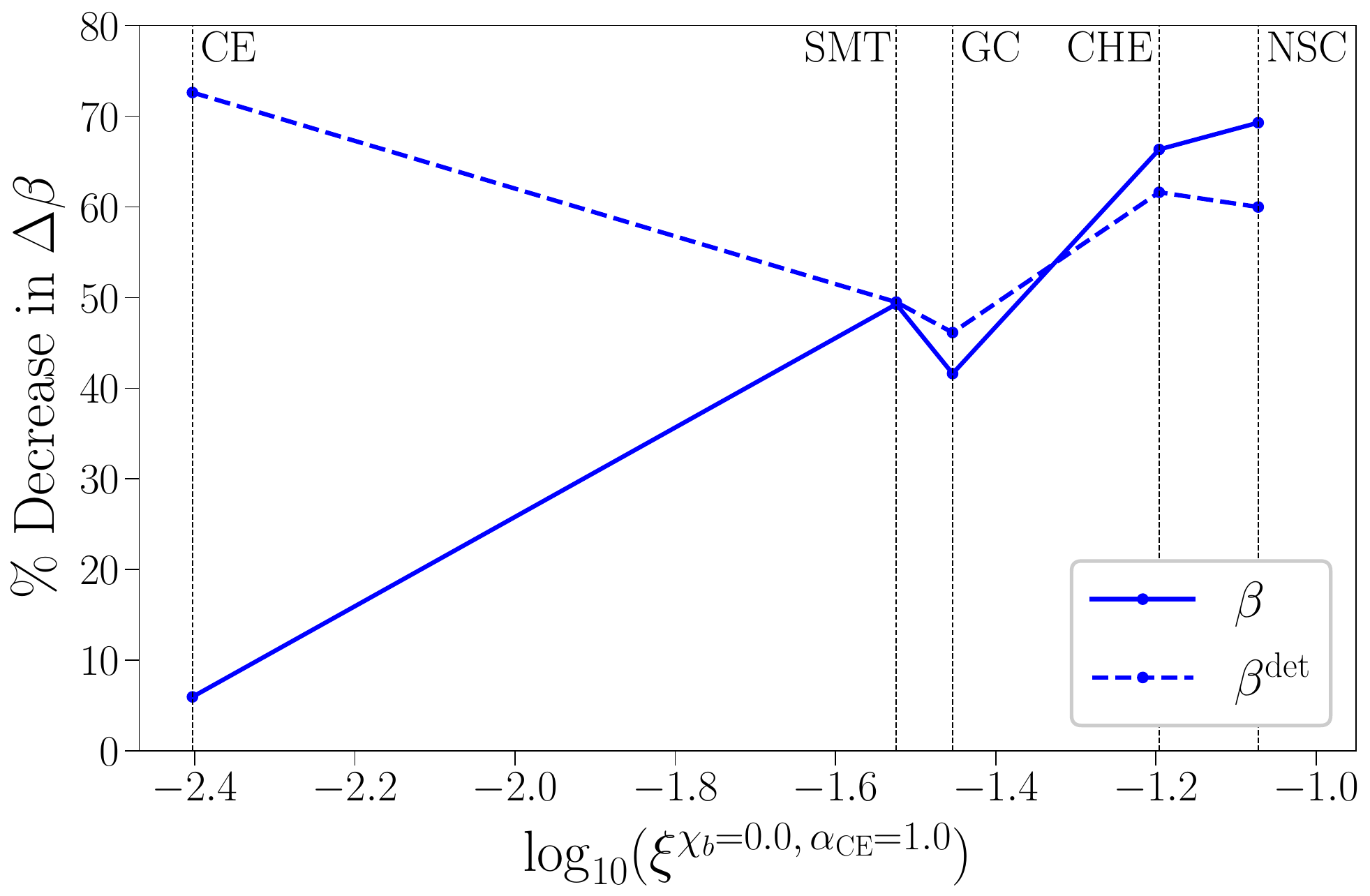}
    \caption{The percent decrease in the branching fraction posterior 90\% symmetric credible interval width from 50 to 250 mock detections plotted against the channel's detection efficiency. 
    The black dotted vertical lines mark which points correspond to which formation channel. 
    While the underlying branching fractions (solid line) converge faster with \Nobs in channels with higher detection efficiencies, no such correlation is seen in the detectable branching fractions (dashed line). }
    \label{fig:alpha1}
\end{figure}

It is worth noting that the scaling of the uncertainty with \Nobs varies significantly between formation channels. 
Some channels have more distinctive features in parameter space (see \cref{fig:popmodels}) that make them easier to distinguish during the inference. 
Additionally, differences in the detection efficiencies of different formation channels likely play a role as well. 
\cref{fig:alpha1} shows the percent decrease in the uncertainty of branching fraction posteriors from $\Nobs=50$ to $\Nobs=250$ against the detection efficiency of its channel $\xi^{\chib, \alphaCE}$ for our fiducial values $\chib=0.0$ and $\alphaCE=1.0$. 
Channels with lower detection efficiencies, most notably the \ac{CE} channel, appear to scale much more poorly with \Nobs than channels with higher detection efficiencies. 
Indeed, the \ac{CE} channel tends to produce lower-mass black holes at higher redshifts (due to typically shorter delay), leading to fewer of them being detectable (see  \cref{fig:popmodels}). 
We emphasize the effect of the low detection efficiency of the \ac{CE} channel: despite the fact that 40\% of the mock underlying population originates from this channel, only 2, 5, and 17 \ac{CE} \acp{BBH} end up in the mock observations, of the 50, 150, and 250 total detections, respectively. 
The \betaCE posterior not only has the largest uncertainty of all formation channels, but also barely decreases in uncertainty as we increase \Nobs. 
If we instead examine how the \textit{detectable} branching fractions scale with $\Nobs$ in \cref{fig:convergence1-det}, we can see that all detectable branching fractions narrow at similar rates as we increase \Nobs from 50 (first row) to 250 (third row). 
Indeed, we can see in \cref{fig:alpha1} that the detectable branching fractions (dotted line) do not exhibit the dependence of the convergence rate on detection efficiency that the underlying branching fractions (solid line) have. 

Already, we can see some of the difficulties that arise from hierarchical Bayesian inference in the face of large measurement uncertainties, selection effects, and degenerate features in parameter space. 
A common theme, we will explore these problems in further detail in Section \ref{sec:biases}.

Finally, we repeat this analysis for a different set of hyperparameters consisting of equal branching fractions between all formation channels ($\beta_j=0.2$ for all $j$), $\chib=0.2$, and $\alphaCE=1.0$ (see the second row of \cref{tab:universes}). 
We find similar results in the convergence with \Nobs. 
Uncertainties in the underlying branching fraction decrease by up to 47\% from $\Nobs=50$ to $\Nobs=250$, and, consistent with our previous results, the uncertainty in \betaCE does not decrease. 
We find also that the shrinking of the detectable branching fraction (\betadet) posterior uncertainty is more consistent across different channels than the underlying branching fraction. 
Increasing \Nobs causes a strong preference for the true value of $\chib=0.2$; while we have $\BF{\chib=0.2}{\chib=0.1}=1.0$ for $\Nobs=50$, there is no posterior support for $\chib \neq 0.2$ at $\Nobs=150$ and $\Nobs=250$. 
Similar to the previous example, there is no strong preference for the true value of \alphaCE, although it is slightly favored with $\BF{\alphaCE=1.0}{\alphaCE=0.5}=1.7$. 
Figures showing the marginalized posteriors on $\beta$ and $\betadet$ for this set of hyperparameters can be found in Appendix \ref{app:additionalfigs}. 

\section{Biases of Population Inference} \label{sec:biases}

Finally, we investigate the biases that may arise when performing hierarchical inference with the methods described above in Section \ref{subsec:inferencemethod}. 
We again refer to \cref{tab:universes} for the chosen true values of the hyperparameters that we present in subsequent sections; we use the same method for hierarchical inference and simulated \ac{BBH} detections as outlined in Sections \ref{subsec:inferencemethod} and \ref{subsec:injsmethod}, respectively. 
In Section \ref{subsec:singlechannel}, we perform hierarchical inference on sources originating exclusively from each individual formation channel and examine differences in the recovered posteriors.
In Section \ref{subsec:spinbias}, we isolate the effect of the natal black hole spin hyperparameter by comparing the posteriors of universes with different choices of \chib. 
In Section \ref{subsec:mixture}, we explore the consequences of doing hierarchical inference with incomplete information (i.e. excluding a channel from the inference) with detections from a mixture of formation channels. 
Finally, we summarize and discuss our results in Section \ref{subsec:biasdiscussion}. 

\subsection{Inference in Single-Channel Dominated Universes} \label{subsec:singlechannel}

\begin{figure*}[ht]
    \includegraphics[width=\textwidth]{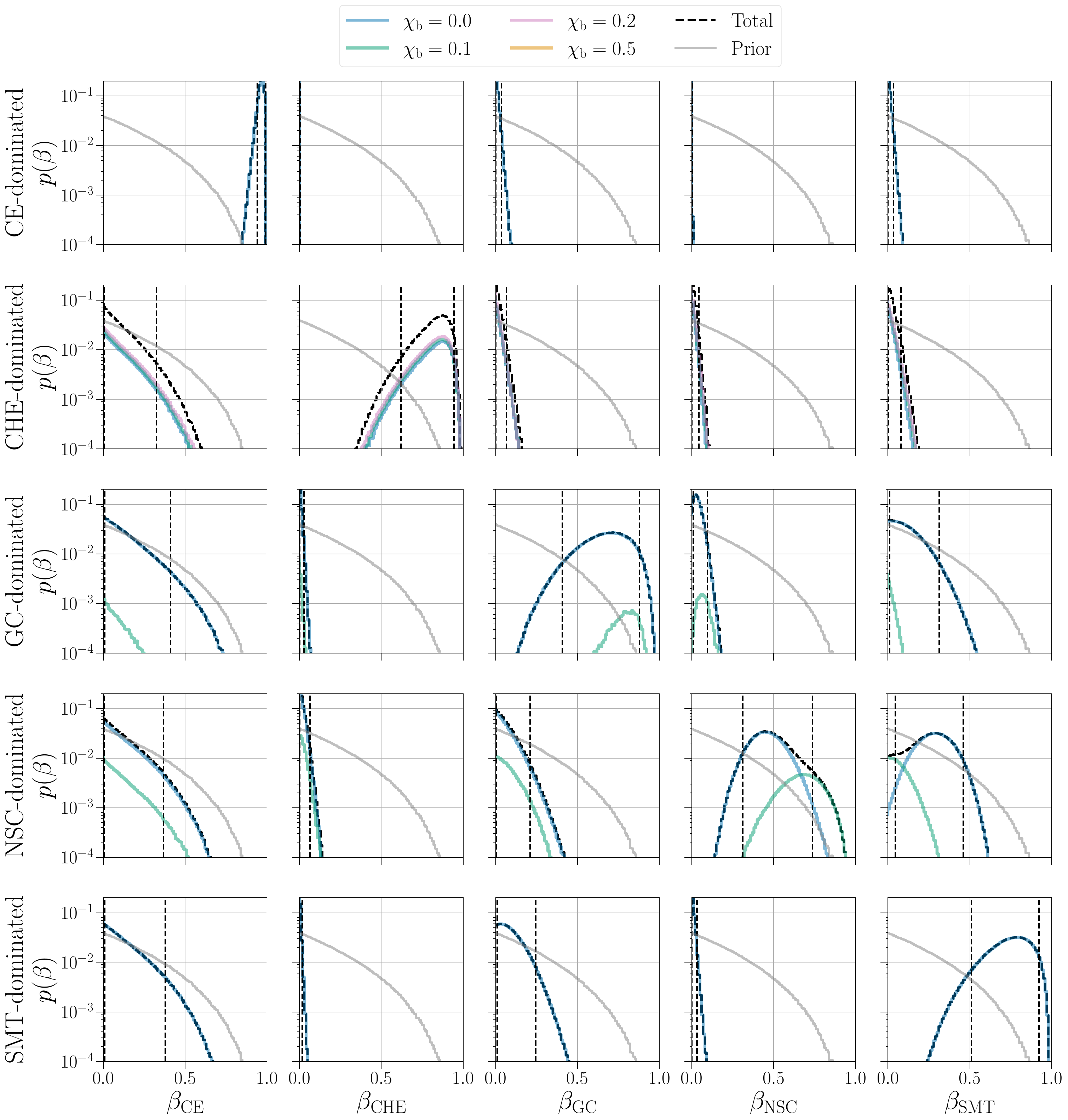}
    \caption{Marginalized branching fraction posteriors inferred from a catalog of 100 mock detections for a universe where the only \ac{BBH} formation channel is (top to bottom) \ac{CE}, \ac{CHE}, \ac{GC}, \ac{NSC}, and \ac{SMT}. 
    As in the top row of \cref{fig:gwtc3}, different colored curves represent support for different values of \chib marginalized over \alphaCE, while the dotted black curve gives the total marginalized posterior. We have also included the prior in gray.
    The vertical lines mark the 90\% symmetric credible interval.}
    \label{fig:singlechannel}
\end{figure*}

For each channel $j$, we perform hierarchical inference on $\Nobs=100$ mock detections in a universe where the entire underlying \ac{BBH} population originates from channel $j$ (i.e. $\beta_j=1$). 
We choose for our true values of the physical prescription $\chib=0.0$ and $\alphaCE=1.0$, although the latter choice only affects the CE-dominated universe. 
Figure \ref{fig:singlechannel} shows the branching fraction posteriors and support for different values of \chib for each single-channel dominated universe. 

In general, none of the branching fraction posteriors have significant support for $\beta=1$, even for the channel that is actually producing the entirety of the \acp{BBH}. 
This happens because we use a flat, symmetric Dirichlet distribution for our prior, which results in a prior preference for a mixture of channels rather than a single dominating channel; this is illustrated by the gray curves in \cref{fig:singlechannel}. 

The 5th percentiles of the dominating-channel branching fraction posteriors for the CE, CHE, GC, NSC, and SMT-dominated universes are $\betaCE=\BetaCESingle$, $\betaCHE=\BetaCHESingle$, $\betaGC=\BetaGCSingle$, $\betaNSC=\BetaNSCSingle$, and $\betaSMT=\BetaSMTSingle$, respectively. 
The degree to which we underestimate the contribution from the dominating channel varies significantly depending on the channel. 
We again highlight the effect of detection efficiency. 
As we saw in Figure \ref{fig:alpha1}, the \ac{CE}, \ac{GC}, and \ac{SMT} channels have the lowest detection efficiencies, especially the \ac{CE} channel. 
Across the different universes under consideration, \betaCE, \betaGC, and \betaSMT (first, middle, and last columns, respectively) have larger uncertainties that compete with and take away from the branching fraction posterior of the dominating channel; because of the lower detection efficiencies of these channels, it is difficult to discern whether this channel is nonexistent or if its contribution to the full set of \textit{detected} observations is minor. 
The case of \betaCE is particularly severe: as it does not strongly deviate from the prior, the 95th percentile for \betaCE is greater than 32\% in all universes for which there is no \ac{CE} contribution to the underlying population. 
The CE-dominated universe (first row) does not suffer from this effect; as a result, only in that universe do we recover with narrow precision that the dominating channel is indeed dominating. 
In the opposite case, the \ac{NSC} channel has the highest detection efficiency. 
In the NSC-dominated universe (fourth row), the median value of the \betaNSC posterior is less than $0.5$, with large contributions to the \ac{BBH} population from the channels with lower detection efficiencies (\betaCE, \betaGC, and \betaSMT). 

There are also several interesting features of the spin model selection. 
First, in the \ac{CHE}-dominated universe (second row), the inference is not able to select the true natal black hole spin of $\chib=0.0$, and instead gives approximately equal support to $\chib=0.0, 0.1$ and $0.2$, as illustrated by the similar heights of the different colored curves. 
Recall that while the aligned-spin \ac{CE} and \ac{SMT} field channels mostly produce \acp{BBH} with $\chieff \approx \chib$ (although \ac{CE} has a tail for higher spins from tidal spin-up) and the isotropic-spin \ac{GC} and \ac{NSC} dynamical channels produce \acp{BBH} scattered around $\chieff \approx 0$, the \ac{CHE} channel uniquely produces \acp{BBH} with $\chieff \gtrsim 0.2$ \textit{irrespective of \chib} due to strong tidal spin-up effects (see \cref{fig:popmodels}). 
As a result, the inference is not able to discern between values of \chib between 0 and the true value 0.2. 
In the universes dominated by dynamical formation channels (\ac{GC}, third row, and \ac{NSC}, fourth row), the selection of the true value of $\chib=0.0$ is less strong, with non-negligible support for $\chib=0.1$ as shown by the green curves. 
In these universes, most detections have $\chieff \approx 0$, and \chib only affects the width of this distribution, a weaker feature to detect. 
Additionally, the sub-population of hierarchical mergers in these channels help to drive the mild support for $\chib > 0$; at lower values of \chib, hierarchical mergers occur more readily due to weaker gravitational recoil kicks. 
Hierarchical mergers can have $|\chieff|$ significantly greater than zero \citep{2005PhRvL..95l1101P,2007PhRvL..98i1101G,2008PhRvD..77b6004B,2020RNAAS...4....2K}, as illustrated by the wings of the marginalized \chieff distribution for the \ac{GC} and \ac{NSC} channels in \cref{fig:popmodels}. 
Because hierarchical mergers form a larger sub-population in \ac{NSC} due to their deeper potential wells and ability to retain post-merger black holes \citep{2009ApJ...692..917M,2009MNRAS.395.2127O,2015MNRAS.448..754H,2016ApJ...831..187A}, this effect is greater for the \ac{NSC}-dominated universe. 
Additionally, in the NSC-dominated universe, the \betaNSC posterior for $\chib=0.1$ (green curve) peaks at a higher value than $\chib=0.0$ (blue curve). 
This is because when $\chib=0.1$, field channels produce fewer of the $\chieff \approx 0$ \acp{BBH} that make up the bulk of the population. 

\begin{figure*}[ht]
    \includegraphics[width=\textwidth]{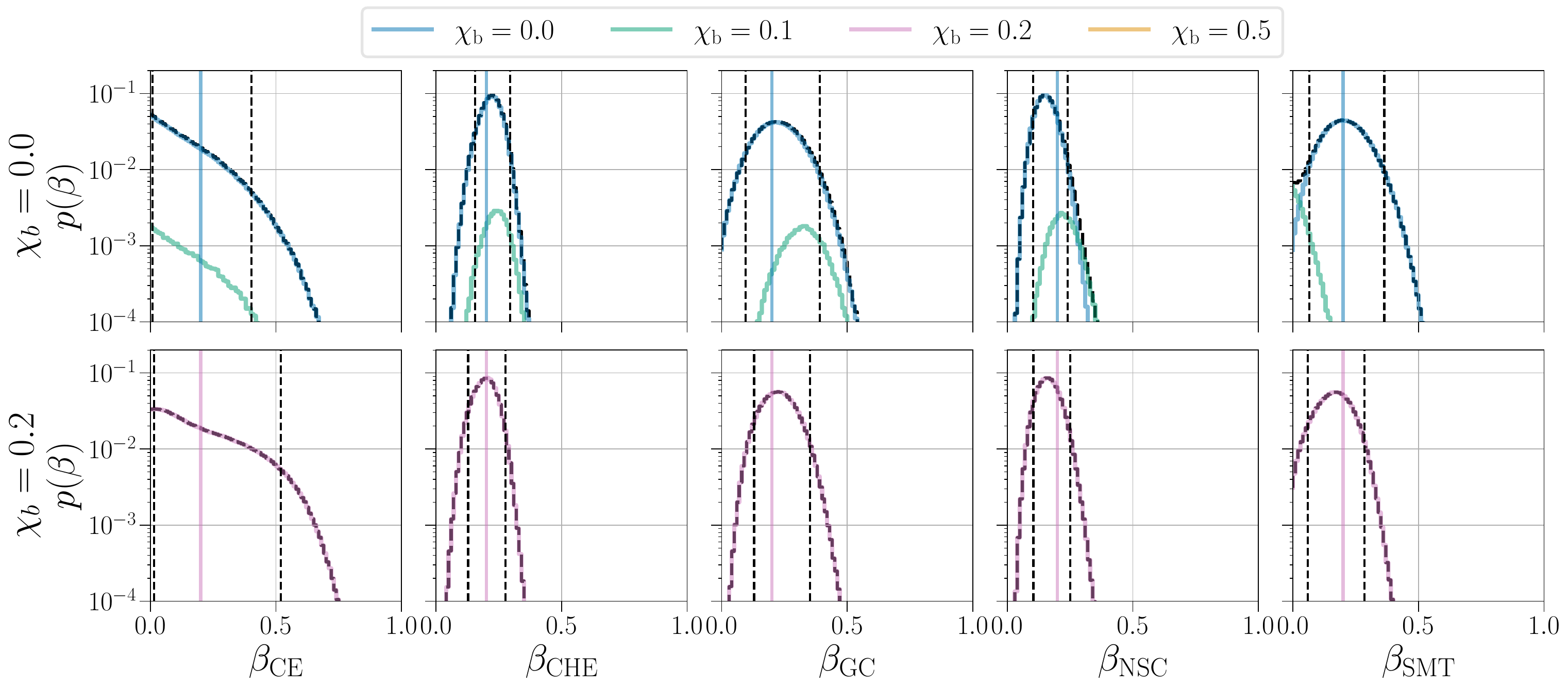}
    \caption{Marginalized branching fraction posteriors in the same format as \cref{fig:singlechannel}, except here the rows show two universes (each with 250 mock observations) with different true values of \chib (0.0, top, and 0.2, bottom). 
    The true values of the branching fractions ($\beta_j=20\%$ for all channels $j$) are given by the solid vertical lines, color-coded by the true value of \chib.}
    \label{fig:spinbias}
\end{figure*}

Finally, although not shown in Figure \ref{fig:singlechannel}, we comment on the inference on $\alphaCE$. 
First, we do favor the true value of $\alphaCE=1.0$ for the \ac{CE}-dominated channel, preferring it over the next most favored model with $\BF{\alphaCE=1.0}{\alphaCE=2.0}=72$, as expected from a universe where all detections are from the \ac{CE} channel. 
We also note consistency with the results of Section \ref{subsec:convergenceresults} in the recovery of \alphaCE for universes where there is no contribution from the \ac{CE} channel; there is a small bias towards higher values of \alphaCE. 
We find $\BF{\alphaCE>1.0}{\alphaCE<1.0}=1.52, 1.54, 1.68,$ and $1.58$ for the \ac{CHE}, \ac{GC}, \ac{NSC}, and \ac{SMT}-dominated universes, respectively\footnote{To be precise, the quantity we calculate is \\ $\left(\BF{\alphaCE=2.0}{\alphaCE=1.0} + \BF{\alphaCE=5.0}{\alphaCE=1.0}\right) / \left(\BF{\alphaCE=0.2}{\alphaCE=1.0} + \BF{\alphaCE=0.5}{\alphaCE=1.0}\right)$}. 

\subsection{Biases in Spin Inference} \label{subsec:spinbias}

To examine the effect of \chib on the hyperposterior, we perform hierarchical inference on different universes with the same underlying branching fractions, but different true values of \chib, using 250 mock detections. 
To isolate the effect of \chib, we choose an equal mixture of formation channels in the underlying population ($\beta_j=0.2$ for all channels $j$). 
We choose $\alphaCE=1.0$. 

\cref{fig:spinbias} shows the marginalized branching fraction posteriors for universes with $\chib=0.0$ and $\chib=0.2$. 
Consistent with our results in Section \ref{subsec:convergenceresults}, we recover the true values of the branching fractions as well as the true value of $\chib$ for both universes. 
Here, we can see that the selection of \chib is non-linear: it is harder to distinguish between lower black hole spins (i.e. $\chib=0.0$ versus $\chib=0.1$) than higher spins. 
While there is no support for other values of \chib in the posterior of the $\chib=0.2$ universe, there is still non-negligible support for $\chib=0.1$ in the $\chib=0.0$ universe, with ${\BF{\chib=0.0}{\chib=0.1}=30}$. 
This is expected; it is difficult to discern between slowly-spinning and non-spinning populations due to the inherent measurement uncertainty of \ac{GW} observations~\citep{2013PhRvD..87b4035B,2014PhRvL.112y1101V,2017PhRvD..95f4053V,2016PhRvD..93h4042P}.

\begin{figure}[ht]
    \includegraphics[width=\columnwidth]{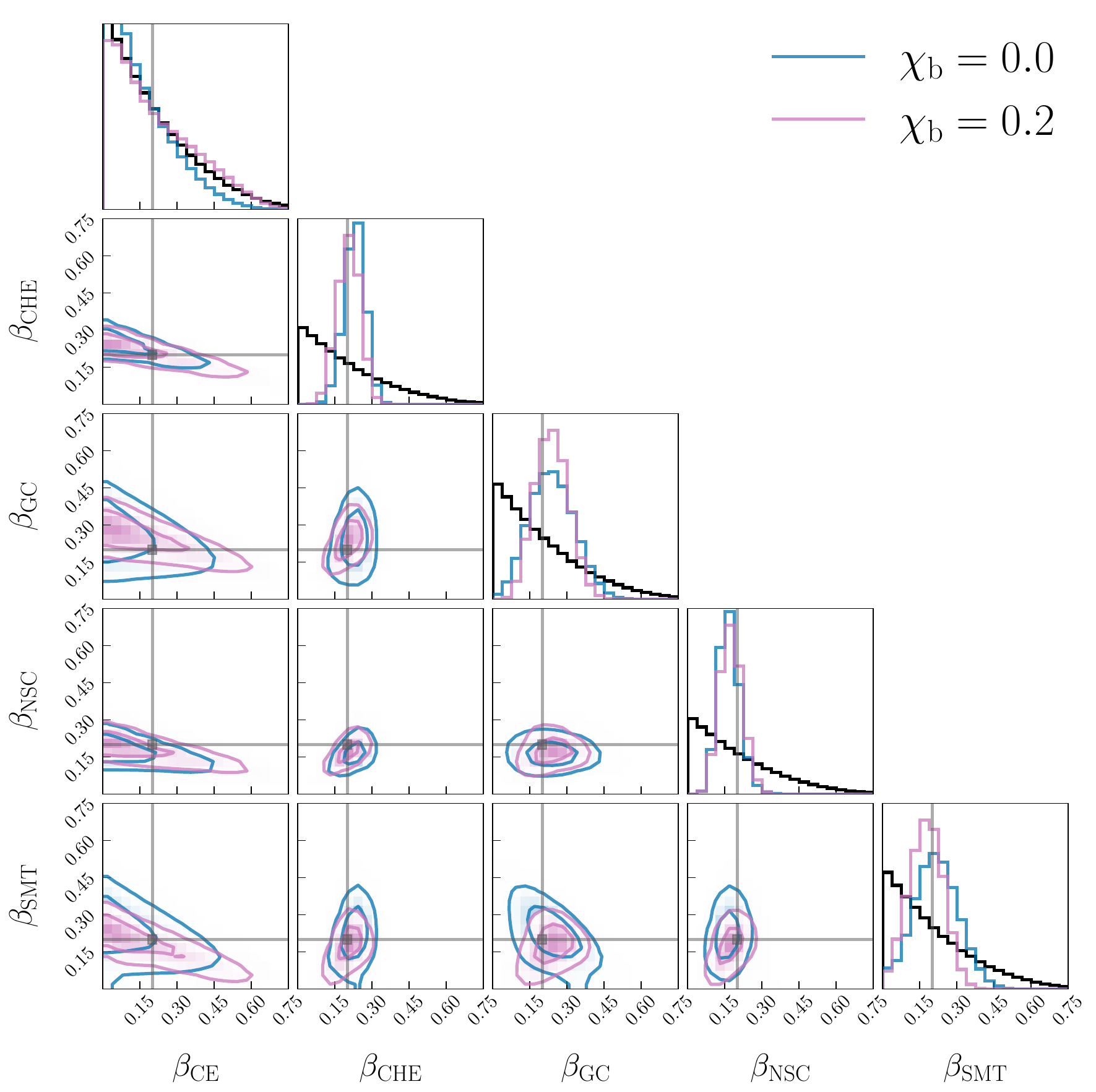}
    \caption{Corner plot of the branching fraction posteriors for an equal mixture universe, with $\alphaCE=1.0$ and $\chib=0.0$ (blue) and 0.2 (pink). 
    Contours correspond to 50\% and 90\% credible intervals. 
    The true values of the branching fractions are marked by the gray lines, while the prior is plotted in black only in the 1D histograms for clarity. }
    \label{fig:corner2}
\end{figure}

Figure \ref{fig:corner2} shows a corner plot of the branching fraction posteriors for both universes. 
Here, we can see the same effect: despite having the same number of mock detections, the data is more informative (yielding a posterior more different from the prior) in the universe with non-zero \chib. 

\subsection{Inference with an Incomplete Set of Populations}\label{subsec:mixture}

\begin{figure*}[p]
    \includegraphics[width=\textwidth]{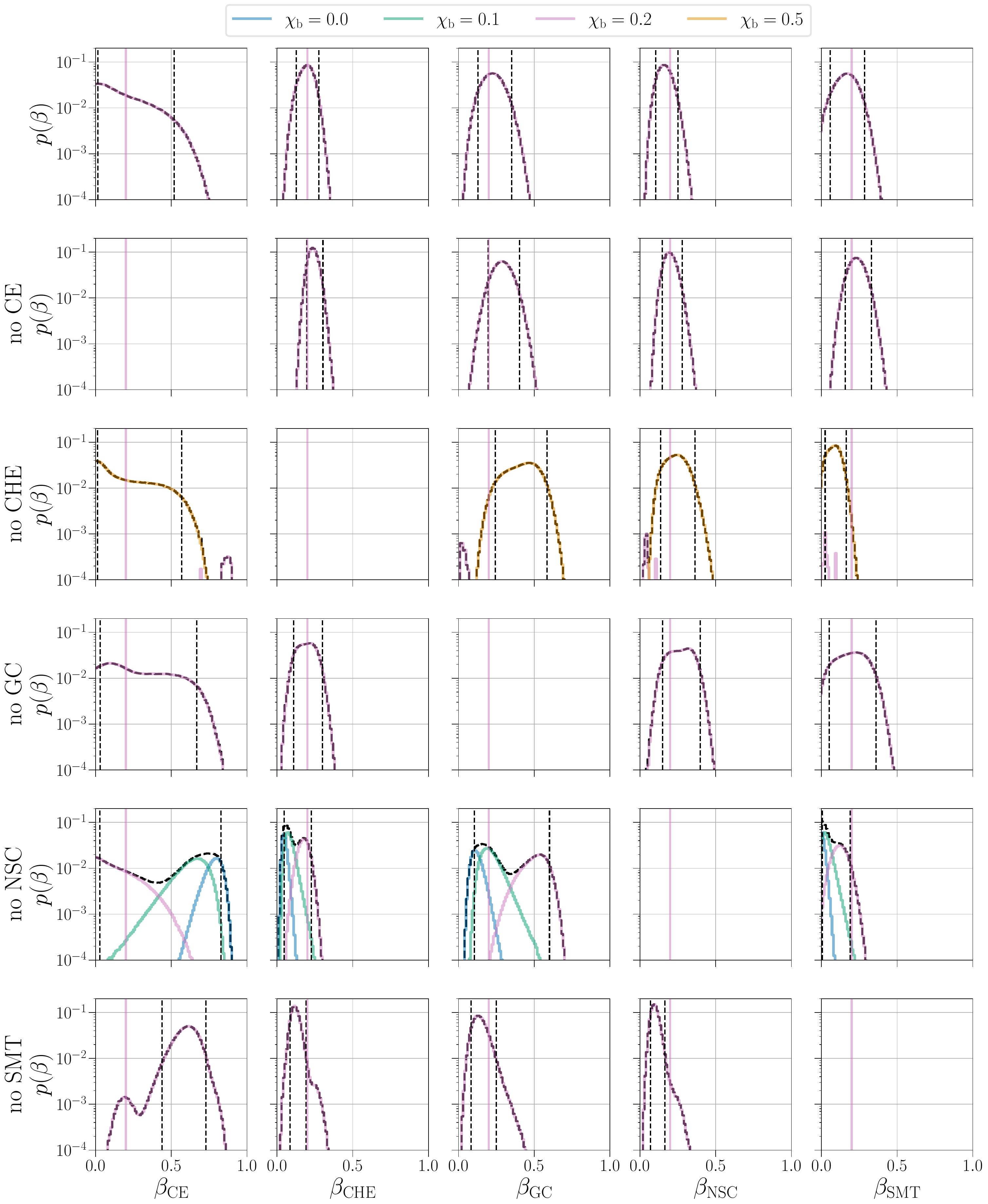}
    \caption{Marginalized branching fraction posteriors. 
    The first row shows the results from the full inference in our $\chib=0.2$, $\alphaCE=1.0$ equal branching-fraction universe with $\Nobs=250$, identical to the second row of \cref{fig:spinbias}, while each of the remaining rows shows the result of inference performed without knowledge of one of the formation channels. 
    The format of this plot is the same as \cref{fig:singlechannel,fig:spinbias}. }
    \label{fig:bias2}
\end{figure*}

Finally, we perform hierarchical inference with one formation channel excluded, such that while all five channels are contributing sources, the inference is only performed with four channels. 
This analysis is motivated by the fact that any population analysis done on real \ac{BBH} data likely does its inference with an incomplete set of formation channels; we most likely do not know the totality of all possible \ac{BBH} formation channels, nor can we model them all accurately and self-consistently. 
We again consider an equal-mixture branching fraction universe ($\beta=0.2$ between all formation channels), and true values for our physical prescription of $\alphaCE=1.0$ and $\chib=0.2$. 
\cref{fig:bias2} shows the marginalized branching fraction posteriors and support for different values of \chib for the full inference as well as for inferences with one channel excluded. 
We defer discussion of biases in \alphaCE selection to Appendix \ref{app:additionalfigs}, and focus on \chib and the formation channel branching fractions in this section. 

By examining which channels receive more or less posterior support as a result of the inference's incomplete knowledge of formation channels, we can infer the correlations between the branching fractions of different formation channels. 
For example, when the \ac{CE} channel is excluded (second row), the increase in branching fraction is spread approximately equally over the other channels, suggesting a negative correlation of $\betaCE$ with the other branching fractions, as is the prior. 
On the other hand, when the \ac{SMT} channel is excluded from the inference (last row), the \betaCE posterior shifts towards higher values, while support for the other channels \textit{decreases}. 

Such correlations are consistent with the branching fraction corner plot from the full inference (\cref{fig:corner2}); the pink contours are relevant to the set of mock detections discussed in this section. 
The uncertainties in \betaCE are the greatest of all channels, and the posterior the least constrained; as such, the marginalized \betaCE posterior closely follows the prior. 
As a result, \betaCE is negatively correlated with all channels, which leads to systematic overestimation when doing inference with a channel excluded. 
In general, we can see that some branching fraction posteriors (i.e. \betaNSC and \betaCHE) are much better constrained than others and are affected the least by the choice of prior. 
As seen in Section \ref{subsec:singlechannel}, the prior can cause a bias towards higher values of $\beta$ for channels with lower detection efficiencies and greater uncertainties. 

Next, we highlight the effect of incomplete formation channel knowledge on the selection of the natal black hole spin \chib. 
We see two cases in which the wrong value of \chib is selected. 
When the \ac{CHE} channel is excluded (third row), we infer a high natal black hole spin of $\chib=0.5$, as indicated by the yellow curve, with a Bayes factor over the true value $\chib=0.2$ of $\BF{\chib=0.5}{\chib=0.2}=373$. 
This is due to the high-\chieff \acp{BBH} that the \ac{CHE} channel produces. 
When the inference does not account for the \ac{CHE} channel, it tries to explain these highly-spinning \ac{CHE} black holes with other field-channel \acp{BBH} spinning at a higher \chib. 
Then, in order to still account for the lower \chieff non-CHE \acp{BBH}, the branching fraction for the \ac{GC} and \ac{SMT} channels are increased and decreased, respectively. 
We remark that no such adjustment is seen for the \ac{NSC} and \ac{CE} channels, despite them having similar features in \chieff space as the \ac{GC} and \ac{SMT} channels, respectively. 

An opposite effect is seen when excluding the \ac{NSC} channel (fifth row), which, as noted above, produces \acp{BBH} scattered around $\chieff=0$ with tails that extend to more positive and negative values due to the presence of hierarchical mergers. 
The inference has constructed two competing explanations in order to explain these lower-spinning \acp{BBH}: low-\chib \ac{CE} \acp{BBH} (represented by the green and blue curves with posterior support for high \betaCE and low \betaGC) and higher-\chib \ac{GC} \acp{BBH} (represented by the pink curve with support for low \betaCE and high \betaGC). 
This highlights the degeneracy between low-spinning field channels and high-spinning dynamical channels in producing similar features in the population \chieff distribution. 
This case is especially remarkable because the green and blue curves ($\chib=0.0$ and $\chib=0.1$) for the \betaCE posterior bears striking resemblance to the \betaCE posterior inferred from GWTC-3 data (\cref{fig:gwtc3}), even though these features are purely an artifact of the inference neglecting a single \ac{BBH} formation channel. 
We again remark on the difference between the \betaCE and \betaSMT posteriors: despite the fact that both \ac{CE} and \ac{SMT} are field channels with the similar features in \chieff space, support for \betaSMT does not increase (and rather decreases) in the low-\chib model as a result of the exclusion of the \ac{NSC} channel. 
One reason why this may occur is that \ac{SMT} \acp{BBH} cannot go through tidal spin-up, unlike \ac{CE} \acp{BBH}, and hence accounts for a narrower range of \chieff concentrated around \chib. 
Therefore, \betaSMT can be strongly affected by the exclusion of a channel with strong features in \chieff space, especially when the wrong model of \chib is inferred.

As mentioned in the previous paragraph, we also find notable that the inference favors two separate competing explanations when \ac{NSC} is excluded. 
The \ac{GC} and \ac{NSC} marginalized \chieff \acp{KDE} are similar due to the isotropy of spin orientations in these two channels; upon examining the marginalized \acp{KDE} of the other three \ac{BBH} parameters (\mchirp, \q, \z), the \ac{GC} channel appears still to have the closest resemblance to the \ac{NSC} channel. 
Despite this, the inference does not simply overcompensate the exclusion of the \ac{NSC} channel by correspondingly increasing \betaGC, suggesting the influence of higher-dimensional features in parameter space. 
On the other hand, no such effect is seen when the \ac{GC} channel is excluded, whose posterior has a simple overcompensation in \betaNSC and \betaCE. 

Indeed, although we have been able to broadly interpret these posteriors by focusing on different channels' features in \chieff space, there must be other subtle effects at play. 
We have shown in multiple ways the differences between the \betaCE and \betaSMT posteriors and the \betaGC and \betaNSC posteriors, despite having similar features in \chieff space. 
Although the exclusion of each channel has its own unique and interesting consequences, there appears to be a bias for the \ac{CE} and \ac{GC} channels, systematically underestimating the \ac{CHE} and \ac{NSC} channels. 
We point again to detection efficiency: of the field and dynamical channels, respectively, the \ac{CE} and \ac{GC} channels have by far the lowest detection efficiencies. 
With the uniform branching fraction spread in our current set of hyperparameters, only 2\% of detections are expected to be from the \ac{CE} channel (versus 30\% and 16\% from the \ac{CHE} and \ac{SMT} channels), and 16\% from the \ac{GC} channel (versus 36\% from the \ac{NSC} channel). 

Finally, we have repeated this analysis for the set of hyperparameters used in the convergence analysis in Section \ref{sec:convergence}, with an unequal mixture of channels and $\chib=0.0$. 
In this universe, the bias towards overestimating \betaGC is more severe due to the lack of spin information from our choice of \chib. 
Consistent with our previous discussion, there again seems to be a bias towards the \ac{CE} and \ac{GC} channels when one channel is excluded. 
Due to the \ac{CHE} and \ac{NSC} channels' high detection efficiencies, the \betaCHE and \betaNSC posteriors support their low true values (5\% for both channels) with small uncertainties. 
The corresponding plot of the branching fraction posteriors for this universe can be found in Appendix \ref{app:additionalfigs}.

\subsection{Discussion} \label{subsec:biasdiscussion}

In this section, we conducted several investigations of different biases that arise in hierarchical Bayesian inference based on astrophysical formation models of \acp{BBH}. 
We summarize the key takeaways as follows:

\begin{enumerate}
    \item Single-channel dominated universes (Section~\ref{subsec:singlechannel})
    \begin{itemize}
        \item Even when our models of the underlying formation channels are perfectly accurate, at $\Nobs=100$ the data are still relatively uninformative due to information loss from parameter estimation and low detection rates. 
        Thus, results of inference are still influenced by the choice of a flat prior and exhibit bias towards a mixture of formation channels. 
        This caveats our result from Section~\ref{subsec:application_to_gwtc3} that no single channel dominates the underlying \ac{BBH} population, from the inference on GWTC-3 data.
        \item It is difficult to precisely infer the true value of \alphaCE, as it only affects the \ac{CE} channel, which produces very few detectable \acp{BBH} due to its low detection efficiency relative to the other channels considered. 
        Only with mock catalogs where the \ac{CE} channel dominates the detections can we recover the true value of \alphaCE, as expected. 
    \end{itemize}
    \item Biases in spin inference (Section~\ref{subsec:spinbias})
    \begin{itemize}
        \item It is easier to infer higher values of \chib than lower values. 
        When \chib is low ($0.0$ or $0.1$), one has less spin information, and uncertainties in both the branching fraction posteriors and the selection of the true value of \chib are greater. 
    \end{itemize}
    \item Inference with incomplete populations (Section~\ref{subsec:mixture})
    \begin{itemize}
        \item Both the branching fraction posteriors and the inferred value of \chib can be heavily affected if the inference is performed without knowledge of all contributing formation channels. 
        Some branching fractions are overestimated or underestimated by a factor of $\sim 3$ or more from the exclusion of a formation channel, and ignoring channels that produce particularly high or low \chieff \acp{BBH} can cause the inference to strongly select an incorrect value of \chib. 
        \item Degeneracies exist between different sets of hyperparameters that can make it difficult for the inference to discriminate between them. 
        For example, it can be difficult to distinguish between field \acp{BBH} with low \chib and dynamical \acp{BBH} with higher \chib, due to \chieff being the only spin information used in the inference (which in turn is due to the fact that \chieff is arguably the only spin parameter that can be measured for all \acp{BBH} with advanced detectors). 
        There are likely more subtle degeneracies and correlations in higher-dimensional parameter space that are difficult to explain from the marginalized distributions, but nonetheless play a role in the inference. 
        \item Inference on the underlying branching fractions can be biased due to the varying detection efficiencies of different channels. 
        In particular, the \ac{CE} channel has a detection efficiency nearly an order of magnitude below the other channels, causing large measurement uncertainties in \betaCE and a relatively uninformed (i.e. close to the prior) posterior for \betaCE. 
        As a result, the exclusion of a channel from the inference usually results in the \betaCE posterior support extending to higher values; similar effects can be seen in other low detection efficiency channels, such as the \ac{GC} and \ac{SMT} channels. 
    \end{itemize}
\end{enumerate} 

\section{Conclusions}\label{sec:conclusions}

Understanding the physical processes and formation environments of compact binary mergers is one of the most pressing questions in \ac{GW} astrophysics. 
In this paper, we pair the most recent catalog of \ac{BBH} mergers provided by the \ac{LVK} with an expansive, self-consistent suite of astrophysical models to investigate the origins of \ac{BBH} mergers. 
Consistent with \cite{2021ApJ...910..152Z}, we find that given our set of astrophysical models, multiple formation channels are likely contributing to the observed population (though see Section~\ref{subsec:singlechannel} for a caveat).
We demonstrate both the predictive power of our inference methodology and its scaling with future detections by generated mock observations with realistic measurement uncertainties from synthetic universes with known branching fractions and physical prescriptions. 
Perhaps most important, we also demonstrate the pitfalls of this type of inference, particularly how an incomplete census of formation models or incorrect physical assumptions can lead to significant biases in inference. 
This work should be treated as a cautionary tale for those attempting to understand relevant physical processes leading to compact binary mergers and formation environments of compact binary progenitors, as inference can be severely compromised if models suffer from inaccuracies of incompleteness. 

Though the suite of \ac{BBH} formation channel models used in this work are state-of-the-art and apply self-consistent physical treatments where possible, they in many ways can be treated as exemplary. 
Given the numerous uncertainties in massive-star evolution, binary physics, compact object formation, and environmental effects, it is currently impossible to construct models with complete physical accuracy or to fully explore all the uncertainties that impact the source property predictions of population synthesis. 
Regardless, the biases demonstrated in this analysis are a generic concern when performing inference based on an incomplete or inaccurate set of astrophysical models. 
We do not suggest that such studies have no utility; compared to population inference that rely on heuristic or flexible models, studies such as these have the benefit of translating directly to physical constraints, albeit requiring proper caveats. 
Despite the potential issues with such analyses, we anticipate that given the diversity of \ac{BBH} properties observed to date, the key result of multiple formation channels contributing to the detected population of \acp{BBH} remains robust. 

A potential concern one might have when considering multiple formation channels for the production of \ac{BBH} mergers is how the universe could conspire to have multiple distinct formation pathways, governed by unique physics, to contribute to the population of merging \acp{BBH} at a similar rate. 
Occam's razor would suggest that this is an unlikely scenario. 
However, astrophysical transients have been shown in many instances not to obey this principle~\citep{2021JAVSO..49...83M}. 
Many channels of \ac{BBH} formation have predicted rates within the same order-of-magnitude~(\citealt{2021RNAAS...5...19R,2022MNRAS.516.5737B}, see \citealt{2022LRR....25....1M} for a review) and the selection effects inherent to \ac{GW} detection are certainly capable of causing sources from intrinsically rare channels to be heavily represented in the detected population. 
Future observations and improved population synthesis routines may help to more robustly disentangle the relative rates of various compact binary formation channels, and thereby have the capability of placing constraints on underlying physical processes. 
Nonetheless, for the time being we show that it is important to consider the potential biases that can accumulate when accounting for an incomplete picture of compact binary mergers in the universe. 

Multiple avenues can be used in tandem with the analyses presented in this work to help expedite the ability of placing robust constraints on compact binary formation channels. 
In addition to analyses of the full population of \ac{BBH} mergers, observational signatures from single events that are unique to one or a subset of formation channels will help to place constraints on the relative contribution of various formation channels~\citep[e.g.,][]{2021ApJ...921L..43Z}. 
Observational constraints from other probes of compact binary formation outside of \ac{GW} astronomy, such as electromagnetic surveys of \ac{BBH} stellar progenitors, astrometric observations of compact object binaries, identification and host association of gamma-ray bursts and kilonovae, and characterization of pulsar binaries in the Milky Way can all help complement and improve constraints that rely solely on \ac{GW} observations. 
Incorporating such information into astrophysical inference will help population analyses using astrophysical simulations remain pertinent and scale with the rapidly-growing catalog of compact binary merger observations. 

The posterior samples in the analyses presented in this work, the code for calculating the prior at the posterior points (see Appendix \ref{app:ptheta}), and all figures, along with additional figures and the accompanying Jupyter notebook, are available on Zenodo \citep{data_release}.

\begin{acknowledgements}
    The authors thank Sylvia Biscoveanu, Tom Callister, Storm Colloms, Amanada Farah, Jack Heinzel, Colm Talbot, and Noah Wolfe for their valuable comments and suggestions. 
    We also thank the anonymous referee for helpful comments on this manuscript.
    A.Q.C. is partially supported by the MIT UROP program. 
    Support for this work and for M.Z. was provided by NASA through the NASA Hubble Fellowship grant HST-HF2-51474.001-A awarded by the Space Telescope Science Institute, which is operated by the Association of Universities for Research in Astronomy, Incorporated, under NASA contract NAS5-26555. 
    S.V. is partially supported by NSF through the award PHY-2045740.
    This material is based upon work supported by NSF's LIGO Laboratory which is a major facility fully funded by the National Science Foundation. 
\end{acknowledgements}

\appendix

\section{Calculating the Prior at the Posterior Points} \label{app:ptheta}

During hierarchical inference, the goal is to calculate the hyperposterior for our hyperparameters $\vec{\Lambda} = [\vec{\beta}, \chib, \alphaCE]$
\begin{equation}
    p(\vec{\Lambda} | \mathbf{x}) = 
    \frac{\pi(\vec{\Lambda}) p(\mathbf{x} | \vec{\Lambda})}{p(\mathbf{x})},
\end{equation}
as given by Bayes' theorem, where $\mathbf{x} = \{\vec{x}_i\}_i^{\Nobs}$ is the set of \ac{BBH} detections and
\begin{equation}
    p(\mathbf{x} | \vec{\Lambda}) = 
    \prod_{i=1}^{N_\mathrm{obs}} \frac{p(\vec{x}_i)}{\int p(\vec{\theta} | \vec{\Lambda})P_\mathrm{det}(\vec{\theta}) \,\mathrm{d}\vec{\theta}} \int
    \frac{p(\vec{\theta}_i | \vec{x}_i) p(\vec{\theta}_i | \vec{\Lambda})}
    {\pi(\vec{\theta}_i)} \,\mathrm{d}\vec{\theta}
\end{equation}
is the hyperlikelihood (see Appendix D of \citealt{2021ApJ...910..152Z}, and \citealt{2015PhRvD..91b3005F,2019MNRAS.486.1086M,2022hgwa.bookE..45V} for reviews). 
Here, we divide out the parameter estimation prior $\pi(\theta)$ evaluated at each point $\vec{\theta_i}$ as we integrate over the space of \ac{BBH} parameters $\vec{\theta}=[\mchirp, \q, \chieff, \z]$. 
We approximate this integral via a Monte-Carlo discrete sum over the posterior samples. 
Therefore, it is necessary to calculate the prior at each posterior sample point. 

\cite{2021ApJ...910..152Z} did this by constructing a 4-dimensional Gaussian-kernel \ac{KDE} with the prior samples in the GWTC-1 and GWTC-2 data releases. 
Due to the potentially prohibitive behavior of high-dimensional \acp{KDE} with insufficient training samples, we choose instead to evaluate the prior at each posterior sample analytically by using the analytical priors from the GWTC-2.1 and GWTC-3 data releases and applying the appropriate Jacobians \citep[see][for a review]{2021arXiv210409508C}. 

\begin{figure*}[t]
    \centering
    \includegraphics[width=\textwidth]{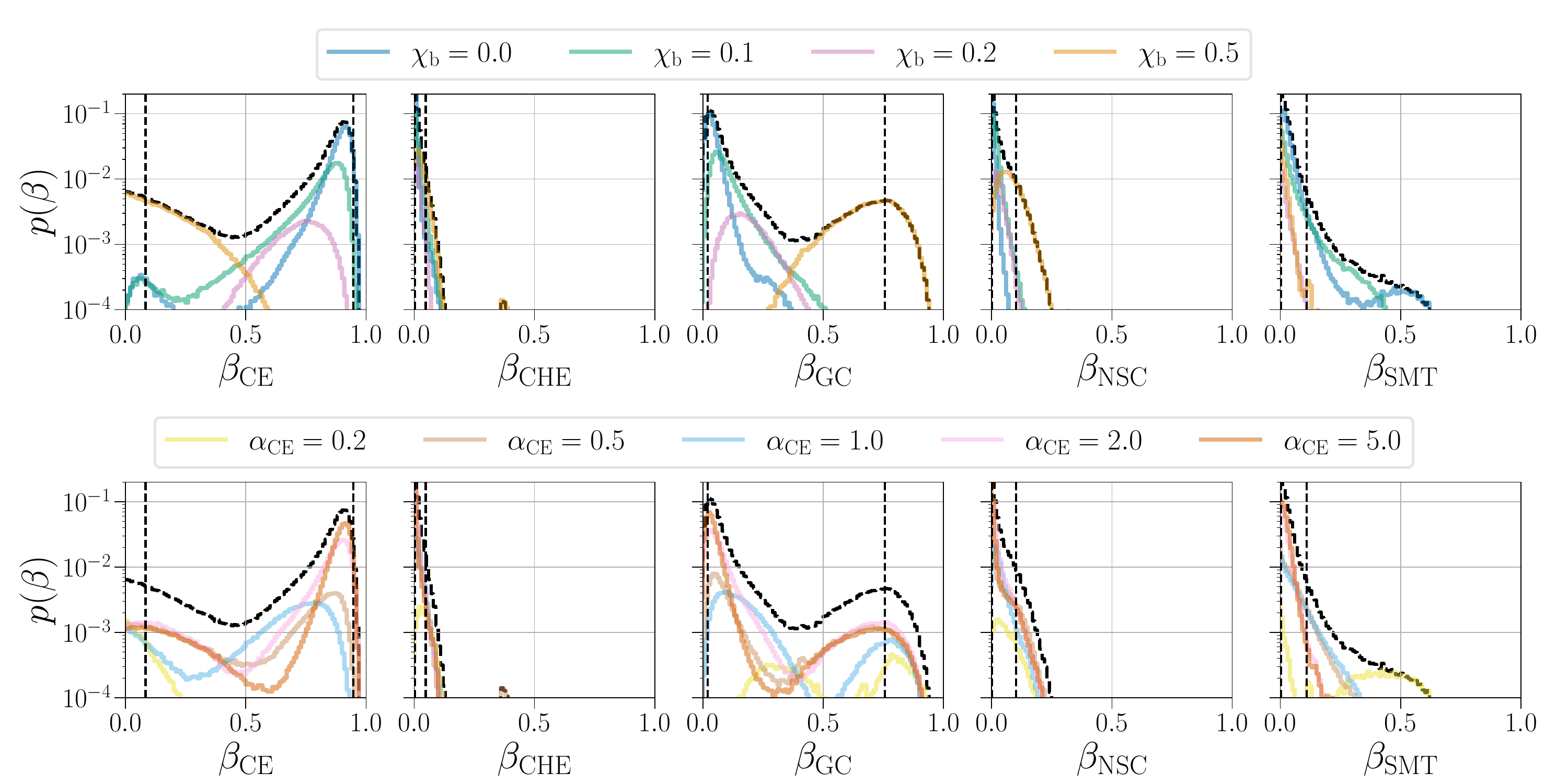}
    \caption{The same as \cref{fig:gwtc3}, showing the marginalized branching fractions inferred from 68 events in the cumulative GWTC-3 catalog, but with the prior at the posterior points evaluated with a \ac{KDE}, rather than analytically.\vspace{3ex}}
    \label{fig:gwtc3-kde}
\end{figure*}

\cref{fig:gwtc3-kde} shows the marginalized branching fraction posteriors inferred from GWTC-2.1 and GWTC-3 data, but with the prior $\pi(\theta_i^k)$ for each event $i$ evaluated at each posterior sample $k$ calculated via a 4-dimensional Gaussian \ac{KDE} constructed from the prior samples provided from \ac{LVK} data releases, as in \cite{2021ApJ...910..152Z}. 
Comparing with \cref{fig:gwtc3}, determining $\pi(\theta_i^k)$ in this way gives rise to some noisy features in the posterior, such as non-trivial support for $\chib=0.2$ and $\chib=0.5$ (pink and yellow curves, respectively). 
While there is no support for $\chib > 0.1$ with analytical evaluation of the prior, with the \ac{KDE} method we have $\BF{\chib \leq 0.1}{\chib > 0.1} = 4.00$. 
Similarly, preference between different values of \alphaCE is also weaker, with $\BF{\alphaCE=5.0}{\alphaCE=1.0}=4.24$, as opposed to $249$. 
The primary notable features in the posterior as discussed in Section \ref{subsec:application_to_gwtc3}, however, still remain robust. 

\FloatBarrier

\section{Additional Figures} \label{app:additionalfigs}

\begin{figure*}[h]
    \centering
    \includegraphics[width=\textwidth]{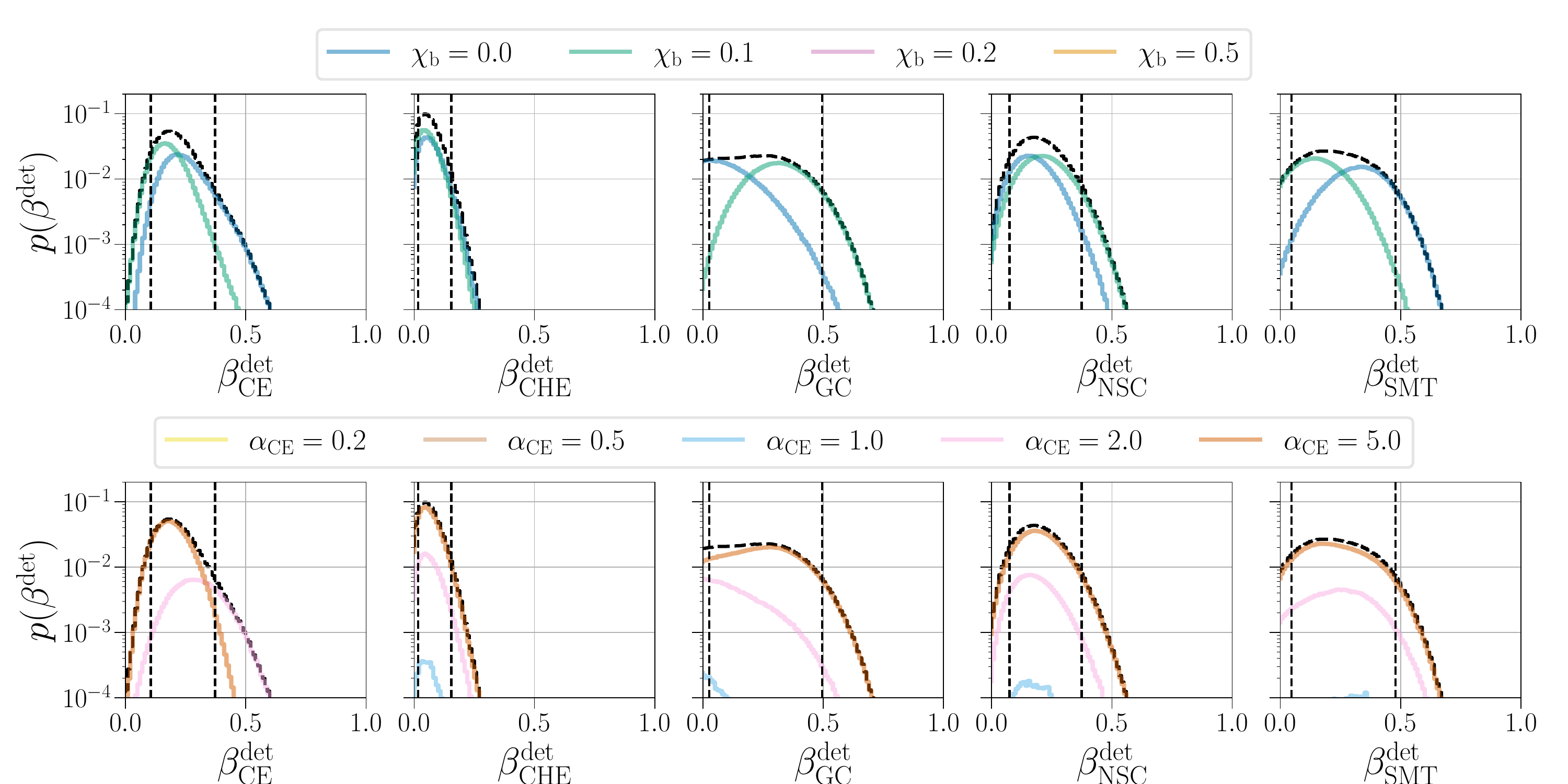}
    \caption{The same as \cref{fig:gwtc3}, except here we show detectable branching fractions inferred from GWTC-2.1 and GWTC-3 data rather than underlying branching fractions. }
    \label{fig:gwtc3-det}
\end{figure*}

In this appendix, we show additional plots (starting from \cref{fig:gwtc3-det}). Details are given in each figure's caption.

\begin{figure*}[h]
    \centering
    \includegraphics[width=0.95\textwidth]{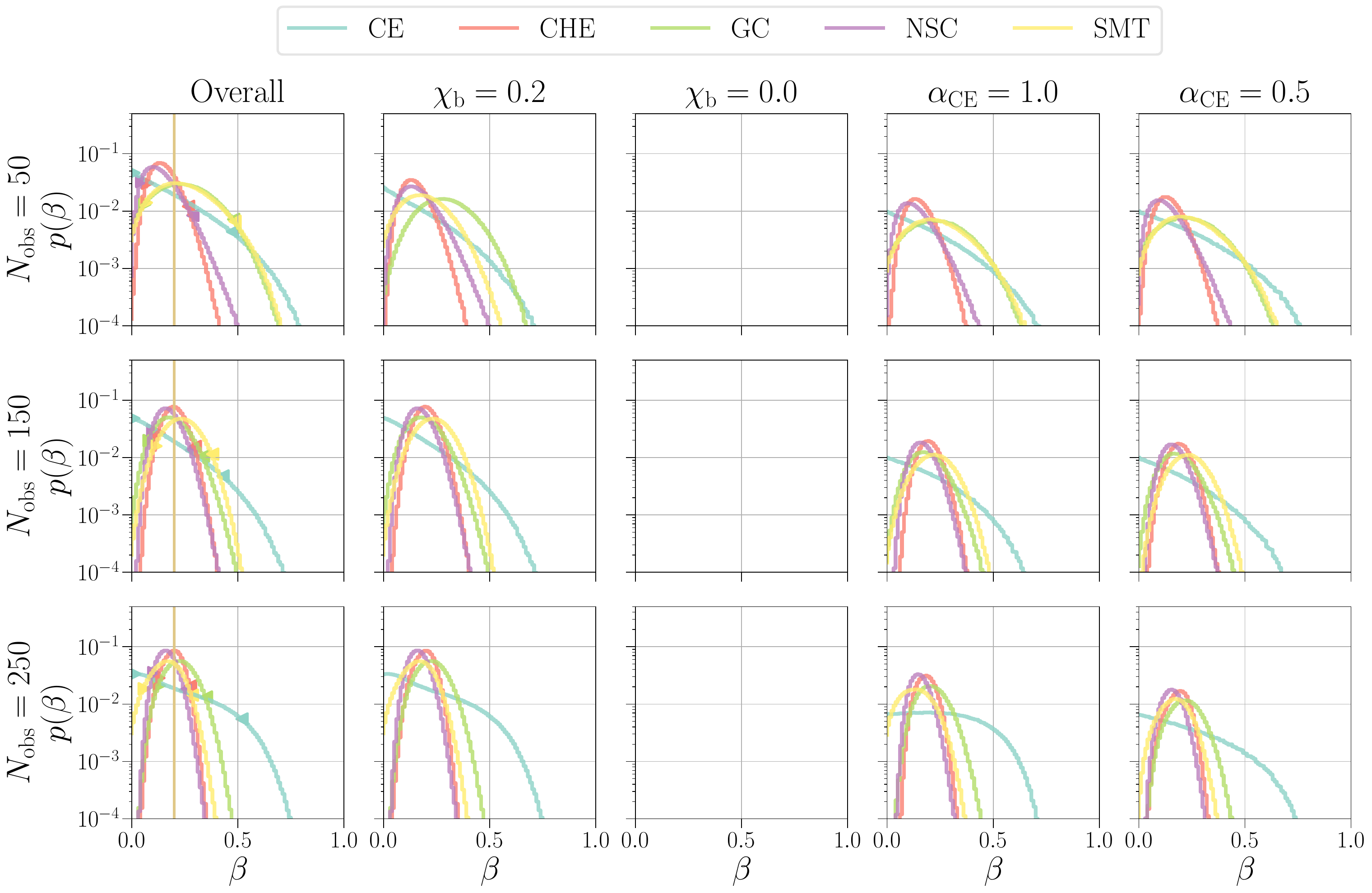} \\
    \includegraphics[width=0.95\textwidth]{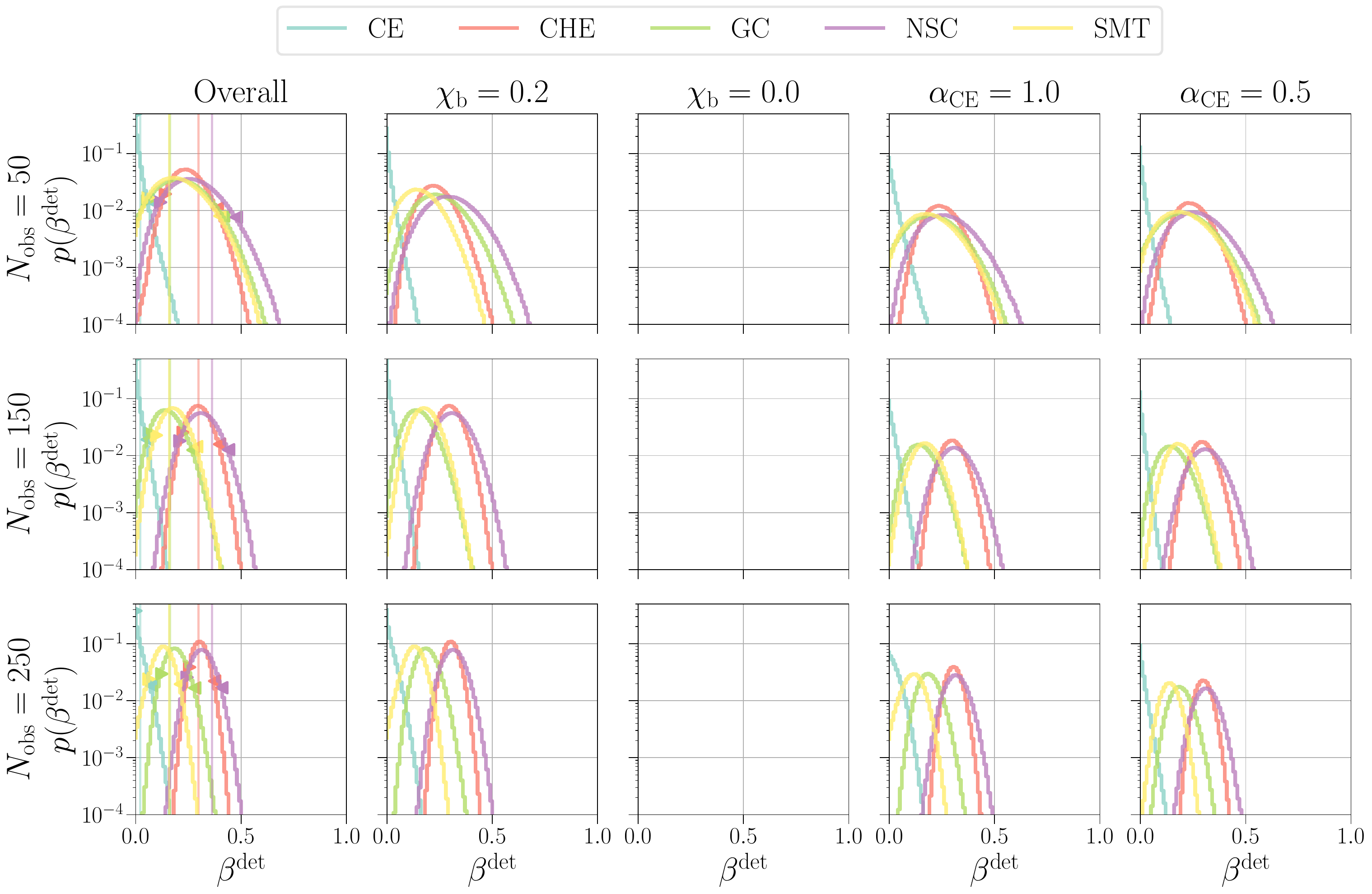}
    \caption{The same as \cref{fig:convergence1} (top, underlying branching fractions) and \cref{fig:convergence1-det} (bottom, detectable branching fractions), showing the convergence of the posteriors with \Nobs, except in this universe we have $\beta=20\%$ across all channels, as well as $\chib=0.2$ and $\alphaCE=1.0$. As in \cref{fig:convergence1,fig:convergence1-det}, the empty plots (middle column) represent the lack of support for the $\chib=0.0$ model, while the populated plots corresponding to $\alphaCE=0.5$ (right column) show that the inference does not significantly favor the true value of $\alphaCE=1.0$.}
    \label{fig:convergence2}
\end{figure*}

\begin{figure*}[p]
    \includegraphics[width=\textwidth]{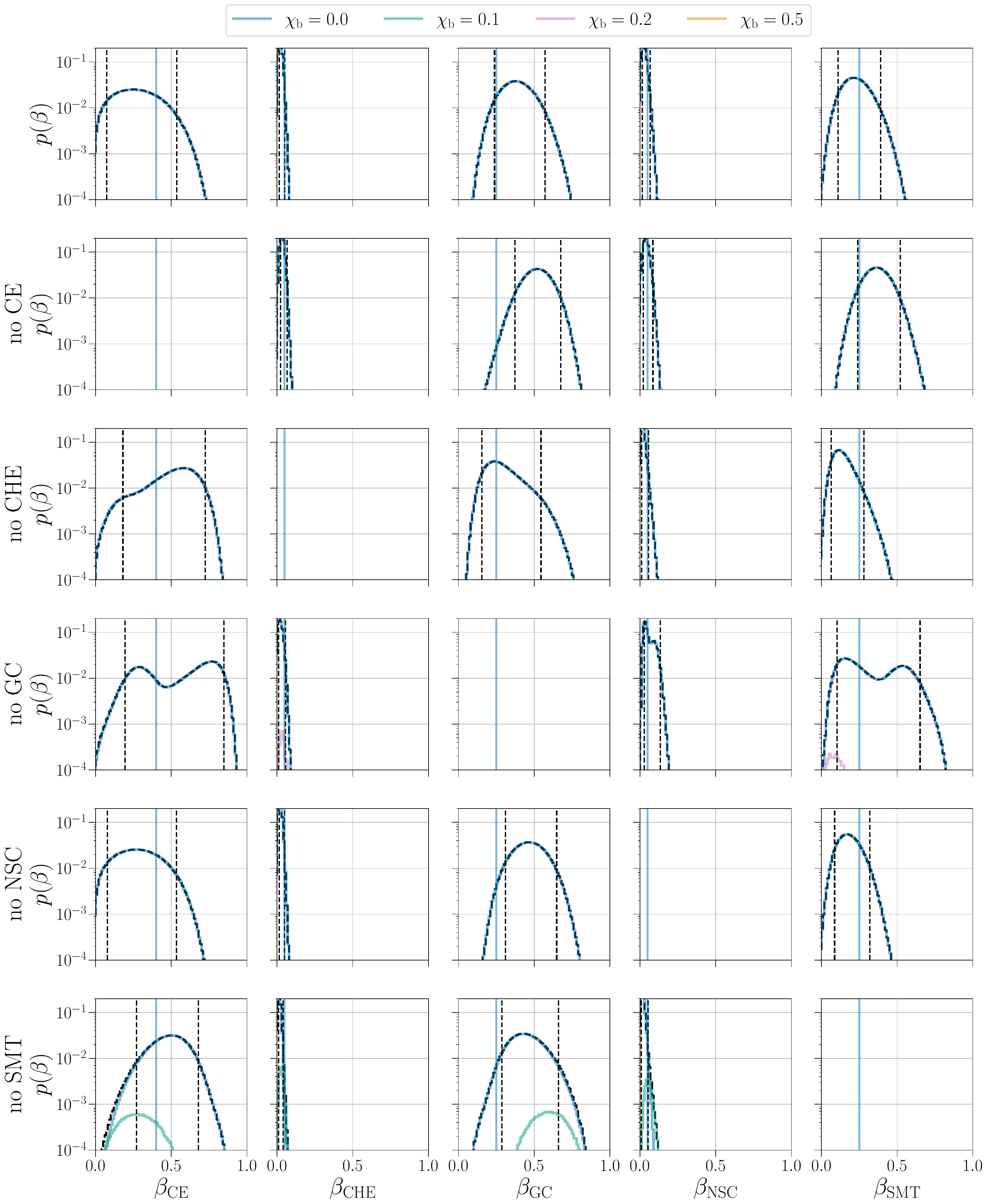}
    \caption{The same as \cref{fig:bias2}, showing the marginalized branching fraction posteriors as a single channel is excluded from the inference, except in this universe we have unequal branching fractions, $\chib=0.0$, and $\alphaCE=1.0$. 
    We also use $\Nobs=250$ for this analysis. }
    \label{fig:bias1}
\end{figure*}

\begin{figure*}[t]
    \includegraphics[width=\textwidth]{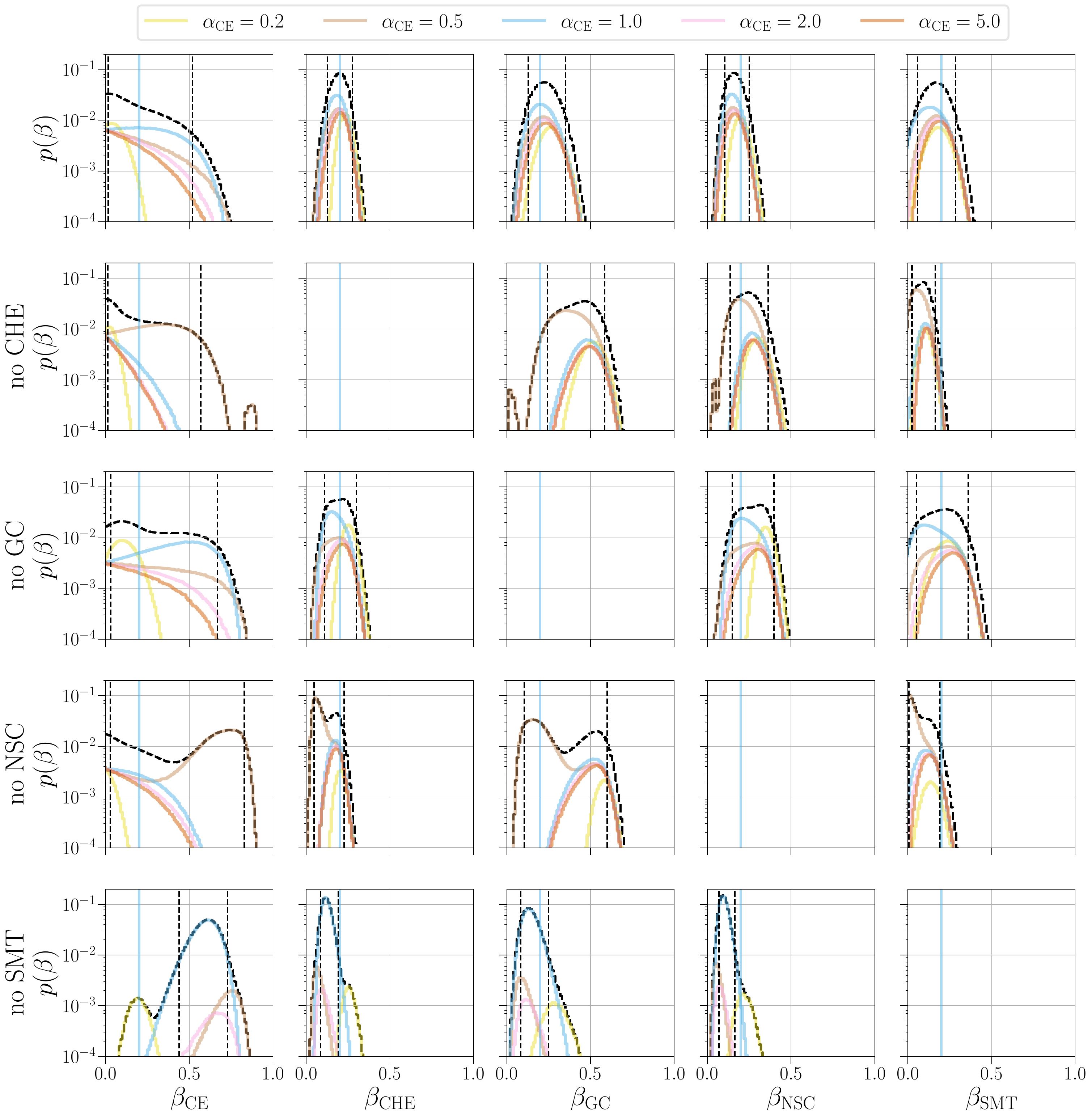}
    \caption{The same as \cref{fig:bias2}, showing the marginalized branching fraction posteriors for each channel in an equal branching fraction universe, except here we show with the colored curves support for different values of \alphaCE, marginalized over \chib. The blue vertical lines show the true values of the branching fractions and of $\alphaCE=1.0$. Note that no value of $\alphaCE$ is inferred when \ac{CE} is not included in the inference, so we do not show it in this plot.
    There are two scenarios in which the wrong value of $\alphaCE$ is strongly preferred: when \ac{CHE} is excluded (second row) and when \ac{NSC} is excluded (fourth row). 
    The selection of a lower value of \alphaCE when \ac{CHE} is excluded is likely due to the necessity of tidally spun-up \ac{CE} \acp{BBH} to explain the high-\chieff \ac{CHE} \acp{BBH}. Additionally, the posteriors of the branching fractions (especially \betaCE) can have varying support depending on the value of \alphaCE. It is possible that the flexibility of the \ac{CE} channel from being able to take on different values of \alphaCE can cause bias towards higher support for \betaCE when marginalizing over \alphaCE, but future investigations involving the exclusion of \alphaCE models from the inference are necessary to confirm this effect.
    }
    \label{fig:bias2alphaCE}
\end{figure*}

\clearpage
\bibliography{export-bibtex,misc}{}
\bibliographystyle{aasjournal}
\end{document}